\RequirePackage{fix-cm}
\documentclass[a4paper]{svjour3}    

\def\makeheadbox{\relax}

\smartqed

\usepackage{graphicx}

\usepackage{latexsym}
\usepackage{ae} 
\usepackage{bm}
\usepackage{eqnarray,amsmath}
\usepackage{booktabs}
\usepackage{amssymb}
\usepackage{amsfonts}
\usepackage{subfigure}
\usepackage{multirow}

\usepackage{hyperref}
\hypersetup{colorlinks,urlcolor=blue}
\usepackage{calrsfs}
\DeclareMathAlphabet{\pazocal}{OMS}{zplm}{m}{n}

\AtBeginDocument{%
  \paperwidth=\dimexpr
    1in + \oddsidemargin
    + \textwidth
    + 1in + \oddsidemargin
  \relax
  \paperheight=\dimexpr
    1in + \topmargin
    + \headheight + \headsep
    + \textheight
    + 1in + \topmargin
  \relax
  \usepackage[pass]{geometry}\relax
}

\renewcommand{\imath}{%
	\ensuremath{\mathrm{i}}}
\newcommand{\rme}{%
	\ensuremath{\mathrm{e}}}
\newcommand{\rmd}{%
	\ensuremath{\mathrm{d}}}

\newcommand{\rvec}{%
	\ensuremath{\mathrm{vec}}}

\newcommand{\tx}{\mathrm{Tx}}
\newcommand{\rx}{\mathrm{Rx}}
\newcommand{\tsup}[1]{\ensuremath{\mathsf{#1}}}

\DeclareMathAlphabet\boldsymbolcal{OMS}{cmsy}{b}{n}

\begin{document}

\hyphenation{non-stationa-ri-ty}
\hyphenation{di-men-sion-al}
\hyphenation{full-di-men-sion-al}

\title{MIMO Channel Reconstruction from Lower Dimensional Multiple Antenna Measurements}

\author{Rimvydas Aleksiejunas}

\institute{
			Telecommunications Research Center, Department of Radiophysics,\\
			Vilnius University, Sauletekio al. 3, LT-10257 Vilnius, Lithuania\\
			\email{rimvydas.aleksiejunas@ff.vu.lt}           
}

\date{}

\maketitle

\begin{abstract}
A method for reconstructing multiple-input multiple-output (MIMO) channel correlation matrices from lower dimensional channel measurements is presented. Exploiting the symmetry of correlation matrix structure enables reproducing higher dimensional MIMO channel matrices from available lower order measurements. This leads to practically important applications allowing prediction of higher dimensional MIMO system capacity. In particular, we study Kronecker-type MIMO channels suitable for reconstructing full channel matrices from partial information about transmit-receive fading in spatial and polarimetric domains and analyze validity conditions for such models. One of the important channel conditions is Doppler frequency related to non-stationarity in the environment. We present simulations of cluster-type scattering model using $2\times2$ MIMO channel correlation matrices to predict performance of $2\times4$ MIMO system including recovery of angular power spectrum. An example of dual circular polarized $2\times4$ MIMO land mobile satellite measurements in 2.5~GHz frequency band illustrates applicability of the method to reconstruct spatial and polarimetric channel correlation matrices for estimating ergodic channel capacity from single-antenna or uni-polarized measurements.
\end{abstract}

\section{Introduction}

MIMO channel correlation analysis is widely used for statistical description of multi-antenna transmission system performance. Exchange of statistical radio channel characteristics between transmitter and receiver allows fast link adaptation and enables high data rate services. Numerous channel prediction algorithms have been constructed in the past to recover incomplete or outdated channel state information at the transmitter to enable efficient MIMO channel transmissions \cite{DuelHallenChPred07}, \cite{PhamChPred11}, \cite{AdeogunChPred15}. Spatial correlation reconstruction is of practical importance also for channel emulation in over-the-air testing of MIMO capable terminals \cite{KyostiEmulatingOTA12}, \cite{FanEmulatingOTA13}, \cite{FanReconstructionOTA13}. Multiple channel characteristics such as power angular spectrum (PAS), Doppler power spectrum or power delay profile are emulated according to given statistical distributions. For instance, having defined PAS $p(\Omega)$ over solid angle $\Omega$, channel correlation coefficients $\rho_{mn}$ between antenna pair $m$ and $n$ separated by distance $d_{mn}$ and transmitting on the wavelength $\lambda$ can be given as
\begin{equation}
  \rho_{mn} = \int \rme^{2\pi j \frac{d_{mn}}{\lambda} \cdot \Omega } p(\Omega) \rmd \Omega,  \label{eq-rho-mn}
\end{equation}
and reconstruction of required $\rho_{mn}$ coefficients becomes an optimization problem. MIMO channel state information prediction at the transmitter from SISO channels may in general use both spatial and temporal correlations \cite{WongSpatioTemporal06}, \cite{LiuSpatioTemporal14}.

Kronecker model has been widely used to reconstruct fading channel correlation matrix from lower order correlation matrices in space, time and frequency selective fading \cite{KermoalKron02}, \cite{XiaoKron304}, \cite{HanKron315}. Kronecker channel correlation structure helps to build efficient covariance estimation methods in high dimensions \cite{WernerCovEst08}, \cite{TsiligkaridisCovEst13}. Another domain for correlation matrix decomposition is polarization, in which case spatial degrees of freedom are separated from polarimetric as shown in \cite{Oestges08}, \cite{ClerckxOestgesBook13} and discussed in a recent review \cite{HePolar16}. The polarimetric properties of correlation matrix are widely studied in relation to land mobile satellite channel modeling \cite{Liolis10}, \cite{King12}, \cite{CheffenaLMSPolar12}.

From the channel correlation matrix decomposition methods cited here, it follows that as far as Kronecker channel matrix decomposition is valid, i.e. fading in space, time, frequency or polarization domains is independent and separable, the full channel covariance matrix can be reconstructed from lower dimensional partial correlation matrices in spatial, temporal, spectral or polarimetric domains. This leads to practically important applications of correlation matrix decomposition when one has MIMO channel measurements for lower order MIMO configuration, but needs to predict higher order MIMO system performance. One of the typical cases is estimation of potential MIMO capacity improvement based on existing MIMO configuration measurements before installing higher order MIMO system.

The present paper addresses the question of predicting MIMO channel capacity from available lower order correlation matrices and  evaluates conditions for such model to be valid. One of the important channel conditions is Doppler frequency related to non-stationarity in the environment. In particular, we study possibility to use $2\times2$ MIMO channel correlation matrices to predict performance of $2\times4$ MIMO systems including polarization domain.

In the following section we present a method of reconstructing MIMO channel correlation matrices for Kronecker-type Gaussian channel model, recover PAS from reconstructed correlation matrix and apply correlation matrix reconstruction to time-variant cluster scattering in high-speed train scenario. In Section \ref{sec-dual-pol}, dual-polarized MIMO channel reconstruction is given with application to measured data of land mobile satellite signals. Finally conclusions are drawn.

\section{Correlation matrix reconstruction for cluster scattering channel models}

\subsection{Kronecker-type correlated Gaussian channel model}

Correlation-based channel model is used to illustrate possibility for reconstructing fourth order receiver channel matrix from second order $2\times2$ MIMO channel measurements, assuming the same correlation properties apply for two-unit as well as four-unit receive antennas. Exploiting the symmetry of correlation matrix elements as represented by Equation (\ref{eq-rho-mn}) enables one to reproduce higher order matrix from lower order correlation matrix by proportionally increasing antenna separation distances $d_{mn}$ between adjacent receive antennas of lower dimensional MIMO system.

In this subsection we analyze antenna correlation performance for Kronecker-type MIMO systems described by the simplest stochastic correlation models, namely, the uniform and exponential. We use uniform channel distribution at transmitter and exponential model at receiver site according to \cite{Loyka01}, \cite{Chizhik03}. Correlation matrix elements for $N_{\tx} \times N_{\rx}$ MIMO system can be expressed as
\begin{equation}
	\left[\mathbf{R}_{\rx}\right]_{mn} = {\rho^{\rx}}^{\left|m-n\right|}, \qquad \left[\mathbf{R}_{\tx}\right]_{mn} = \begin{cases}
    1, & \text{for } m = n \\
    \rho^{\tx}, & \text{for } m \ne n
  \end{cases}  \label{eq-exp-uniform}
\end{equation}
where $m, n = 1, \ldots, N_{\rx}$ for receiver and $m, n = 1, \ldots, N_{\tx}$ for transmitter, while $\rho^{\rx}$ and $\rho^{\tx}$ indicate correlation coefficient between two adjacent antenna elements at receiver and transmitter, respectively, such that $\left|\rho^{\rx}\right| \le 1$ and $\left|\rho^{\tx}\right| \le 1$. Here and in the following the notation $\left[ \, \cdot \, \right]_{m n}$ denotes matrix element with indexes $m$ and $n$.

After generating transmitter (Tx) and receiver (Rx) correlation matrices, $\mathbf{R}_{\tx}$ and $\mathbf{R}_{\rx}$, we will find reconstructed Kronecker-type full channel correlation matrix and compare it with the original full-dimension correlation matrices. We start with a Kronecker correlated Gaussian $2 \times 4$ MIMO channel described by correlation matrix $\mathbf{R}_{\mathrm{H}} = \mathbf{R}_{\rx} \otimes \mathbf{R}_{\tx}$. Such system is represented by rank 2 matrix and we wish to show possibility to reproduce channel behavior by a set of partial $2 \times 2$ MIMO systems, which would exhibit the same multiplexing capability as the full $2 \times 4$ system. Original channel matrix $\mathbf{H}$ is built using Cholesky factorization \cite{GolubMatrix96} of circularly symmetric complex Gaussian random matrix with zero mean and unit variance $\mathbf{H}_{\mathrm{iid}} \sim \pazocal{CN}(0,1)$ as
\begin{equation}
\rvec(\mathbf{H}) = \mathbf{R}_{\mathrm{H}}^{1/2} \rvec\left(\mathbf{H}_{\mathrm{iid}}\right), \label{eq-chol-fact}
\end{equation}
where $\rvec(\cdot)$ denotes column-wise vectorization operator and $(\cdot)^{1/2}$ is the matrix square root implemented using Cholesky factorization. Assuming that all antennas are identical and disregarding mutual coupling effects (which is valid for uniform and exponential spatial correlation models) leads to the following channel covariance matrix structure
\begin{equation}
\mathbf{R}_{\mathrm{H}} = \left[ \begin{array}{cccc}
 1       &  \rho^{\rx}_{12} & \rho^{\rx}_{13} & \rho^{\rx}_{14} \\
 {\rho^{\rx}_{12}}^{*} &  1   &  \rho^{\rx}_{12} & \rho^{\rx}_{13} \\
{\rho^{\rx}_{13}}^{*} &  {\rho^{\rx}_{12}}^{*} &  1  & \rho^{\rx}_{12} \\
{\rho^{\rx}_{14}}^{*} & {\rho^{\rx}_{13}}^{*} &  {\rho^{\rx}_{12}}^{*} & 1
\end{array} \right]  \otimes \mathbf{R}_{\tx}, \label{eq-rx-corr4}
\end{equation}
described by elements $\rho^{\rx}_{12}$, $\rho^{\rx}_{13}$, $\rho^{\rx}_{14}$ and their complex conjugates denoted by $(\, \cdot \, )^{*}$. Such covariance matrix structure allows us to express Rx correlation properties using reduced dimensionality covariance matrices, i.e. three distinct covariance matrices of partial $2 \times 2$ MIMO subsystems:
\begin{equation}
\mathbf{R}_{\mathrm{H}_{12}} = \left[ \begin{array}{cc}
 1       &  \rho^{\rx}_{12} \\
 \rho^{\rx^{*}}_{12} & 1
 \end{array} \right]  \otimes \mathbf{R}_{\tx}, \quad
\mathbf{R}_{\mathrm{H}_{13}} = \left[ \begin{array}{cc}
 1       &  \rho^{\rx}_{13} \\
 \rho^{\rx^{*}}_{13} & 1
 \end{array} \right]  \otimes \mathbf{R}_{\tx}, \quad
\mathbf{R}_{\mathrm{H}_{14}} = \left[ \begin{array}{cc}
 1       &  \rho^{\rx}_{14} \\
 \rho^{\rx^{*}}_{14} & 1
 \end{array} \right] \otimes \mathbf{R}_{\tx}. \label{eq-sub-cor}
\end{equation}
Application of Cholesky factorization (\ref{eq-chol-fact}) to partial covariance matrices (\ref{eq-sub-cor}) generates independent snapshots of $2\times 2$ MIMO channel realizations for each possible Rx antenna pair: $\mathbf{H}_{12}$ channel matrix representing correlation properties between Rx antennas \#1 and \#2, $\mathbf{H}_{13}$ representing antennas \#1 and \#3, while $\mathbf{H}_{14}$ -- antennas \#1 and \#4. Note that the same unchanged two-antenna configuration is retained at the transmitter site with correlation properties expressed by matrix $\mathbf{R}_{\tx}$.

The full-dimensional $2\times 4$ reconstructed channel matrix $\hat{\mathbf{H}}$ can be obtained by adding all possible combinations of zero-padded partial channel matrices $\mathbf{H}_{12}$, $\mathbf{H}_{13}$ and $\mathbf{H}_{14}$ in the following way:
\begin{equation}
  \hat{\mathbf{H}} = \left[\begin{array}{c}
    \multirow{2}{*}{$\mathbf{H}_{12}$} \\
     \\
    \multirow{2}{*}{$\mathbf{0}_{2 \times 2}$} \\
     \\
  \end{array}\right] +
  \left[\begin{array}{c}
    \multirow{2}{*}{$\mathbf{0}_{2 \times 2}$} \\
     \\
    \multirow{2}{*}{$\mathbf{H}_{12}$} \\
     \\
  \end{array}\right], \;
  \left[\begin{array}{c}
    \mathbf{0}_{1 \times 2} \\
    \multirow{2}{*}{$\mathbf{H}_{12}$} \\
     \\
     \mathbf{0}_{1 \times 2}
  \end{array}\right] +
  \left[\begin{array}{c}
    \left[\mathbf{H}_{14}\right]_{1\,:} \\
    \multirow{2}{*}{$\mathbf{0}_{2 \times 2}$} \\
     \\
    \left[\mathbf{H}_{14}\right]_{2\,:}
  \end{array}\right], \;
  \left[\begin{array}{c}
    \left[\mathbf{H}_{13}\right]_{1\,:} \\
    \mathbf{0}_{1 \times 2} \\
    \left[\mathbf{H}_{13}\right]_{2\,:} \\
    \mathbf{0}_{1 \times 2}
  \end{array}\right] +
  \left[\begin{array}{c}
    \mathbf{0}_{1 \times 2} \\
    \left[\mathbf{H}_{13}\right]_{1\,:} \\
    \mathbf{0}_{1 \times 2} \\
    \left[\mathbf{H}_{13}\right]_{2\,:}
  \end{array}\right].
  \end{equation}
Here the colons in matrix element indexes $\left[ \, \cdot \, \right]_{m\,:}$ represent all matrix row elements for specified row $m$ and comas between matrices indicate consecutive channel snapshots after pairwise addition of zero padded matrices. $\mathbf{0}_{M\times N}$ denotes zero valued matrices comprising $M$ rows and $N$ columns. In this case the total number of channel realizations increase threefold as a result of these combinations. Such kind of channel matrix reconstruction distorts in some degree the covariance matrix, but leaves unchanged channel matrix statistics from which reconstructed MIMO channel capacity can be estimated.

The difference between reconstructed $\hat{\mathbf{R}}$ and original $\mathbf{R}$ covariance matrices is estimated using relative matrix error $\epsilon$, correlation matrix distance (CMD) $d_{\mathrm{corr}}$ and correlation matrix collinearity (CMC) $c$ defined as \cite{HerdinCMD05}, \cite{GolubMatrix96}
\begin{align}
\epsilon \left(\hat{\mathbf{R}}, \mathbf{R}\right) &= \frac{\| \hat{\mathbf{R}} - \mathbf{R}\|_{\mathrm{F}}}{\sqrt{\| \hat{\mathbf{R}} \|_{\mathrm{F}} \| \mathbf{R} \|_{\mathrm{F}}}}, \label{eq-rel-err}\\
d_{\mathrm{corr}}\left(\hat{\mathbf{R}}, \mathbf{R}\right) &= 1 - \frac{\mathrm{tr}\left\{\hat{\mathbf{R}}\mathbf{R}^{\tsup{H}}\right\}}{\| \hat{\mathbf{R}} \|_{\mathrm{F}} \| \mathbf{R} \|_{\mathrm{F}}}, \label{eq-cmd} \\
c\left(\hat{\mathbf{R}}, \mathbf{R}\right) &= \frac{\left|\mathrm{tr}\left\{\hat{\mathbf{R}}\mathbf{R}^{\tsup{H}}\right\}\right|}{\| \hat{\mathbf{R}} \|_{\mathrm{F}} \| \mathbf{R} \|_{\mathrm{F}}}, \label{eq-cmc}
\end{align}
where $\|\,\cdot\,\|_{\mathrm{F}}$ denotes the Frobenius norm and $\mathrm{tr}\left\{\,\cdot\,\right\}$ is the trace of a square matrix. The similarity of the matrices are identified by $\epsilon$ and CMD values approaching 0 and at the same time CMC being close to 1.

The results of correlation matrix estimation for $2 \times 4$ MIMO configuration with exponential Rx matrix model and uniform Tx side fading according to expressions (\ref{eq-exp-uniform}) is given in Table \ref{tab-kron-corr}. Here the adjacent Tx antenna correlation coefficient is set to $\rho^{\tx} = 0.2$ and $\rho^{\rx}$ is varied between 0.2 and 0.8 with snapshot size of channel realization being $10^{6}$ points. It is evident that increasing $\rho^{\rx}$, accuracy of reconstructing Rx covariance matrix decreases.

\begin{table}[h!]
  \caption{Difference between original $\mathbf{R}_{\rx}$ and reconstructed $\hat{\mathbf{R}}_{\rx}$ receiver correlation matrices for Kronecker type Gaussian channel.}
  \label{tab-kron-corr}
  \begin{tabular}{cccc}
    \toprule
    $\rho^{\rx}$ & $\epsilon \left(\hat{\mathbf{R}}, \mathbf{R}\right)$ & $d_{\mathrm{corr}}\left(\hat{\mathbf{R}}, \mathbf{R}\right)$ &  $c\left(\hat{\mathbf{R}}, \mathbf{R}\right)$ \\
    \midrule
      0.2   &   0.285   &   0.040   &   0.960\\
      0.4   &   0.484   &   0.112   &   0.888\\
      0.6   &   0.645   &   0.191   &   0.809\\
      0.8   &   0.762   &   0.253   &   0.747\\
    \bottomrule
  \end{tabular}
\end{table}

For the same radio channel realizations we calculate cumulative statistical distributions of singular values $\sigma_{i}(\mathbf{H})$, $i = 1, \ldots, \mathrm{rank}(\mathbf{H})$, channel condition numbers $\kappa(\mathbf{H})$ and ergodic capacities $C(\mathbf{H})$ of complex-valued channel matrix $\mathbf{H}$ with dimensions $N_{\tx} \times N_{\rx}$:
\begin{align}
  \kappa\left(\mathbf{H}\right) &= \frac{\sigma_{\max}(\mathbf{H})}{\sigma_{\min}(\mathbf{H})},\\
  C\left(\mathbf{H}\right) &= \mathbb{E}\left\{\log_{2}\det \left[\mathbf{I}_{N_{\rx}} + \frac{\gamma}{N_{\tx}} \mathbf{H}\mathbf{H}^{\tsup{H}} \right] \right\},
\end{align}
where $\sigma_{\max}(\mathbf{H})$ and $\sigma_{\min}(\mathbf{H})$ are, respectively, the largest and smallest singular value in channel matrix $\mathbf{H}$ and $\gamma$ is the signal to noise ratio (SNR).

\begin{figure*}
\centering \subfigure[]{
   \includegraphics[width=0.45\textwidth]{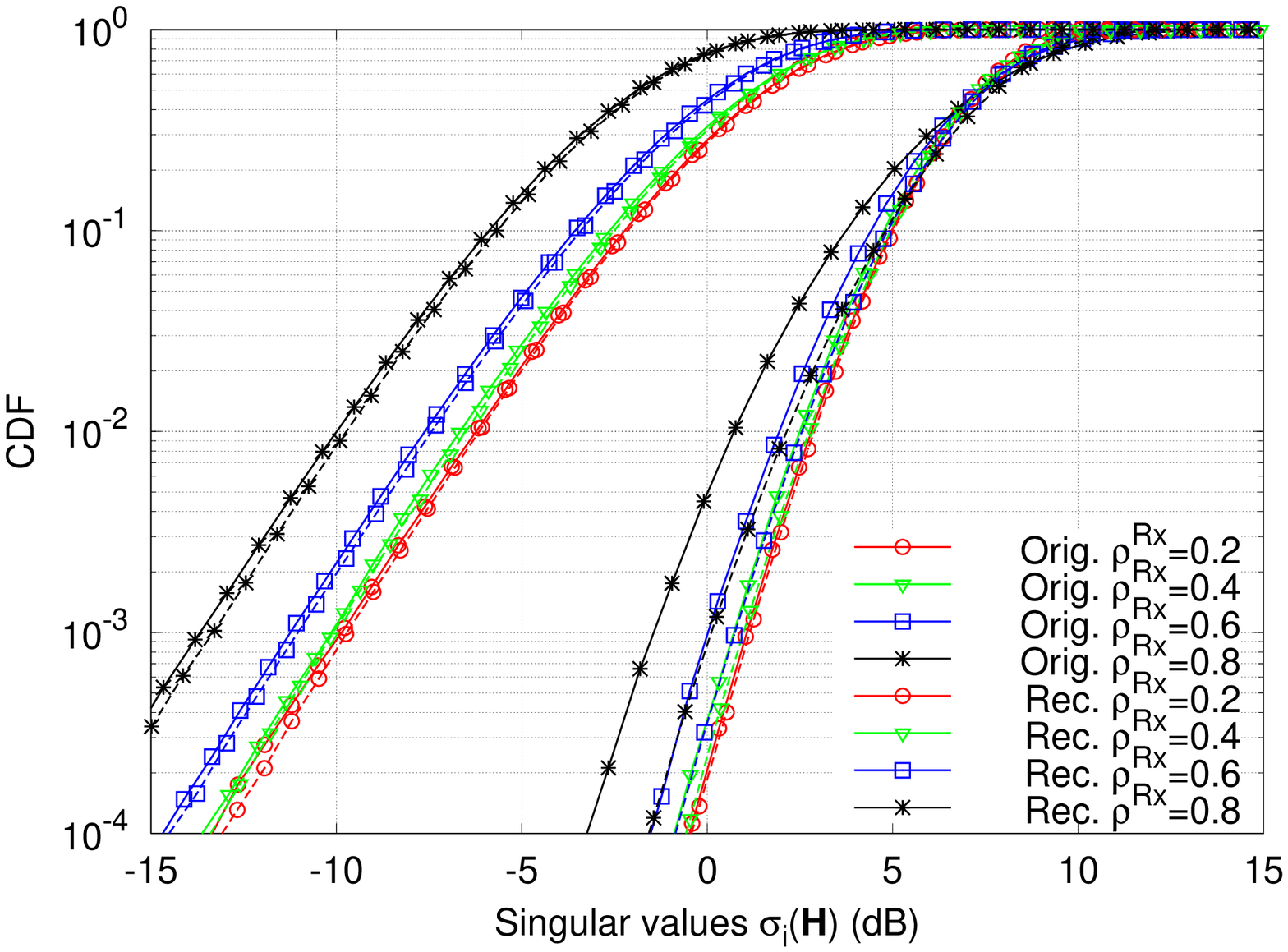}}
\subfigure[]{
   \includegraphics[width=0.45\textwidth]{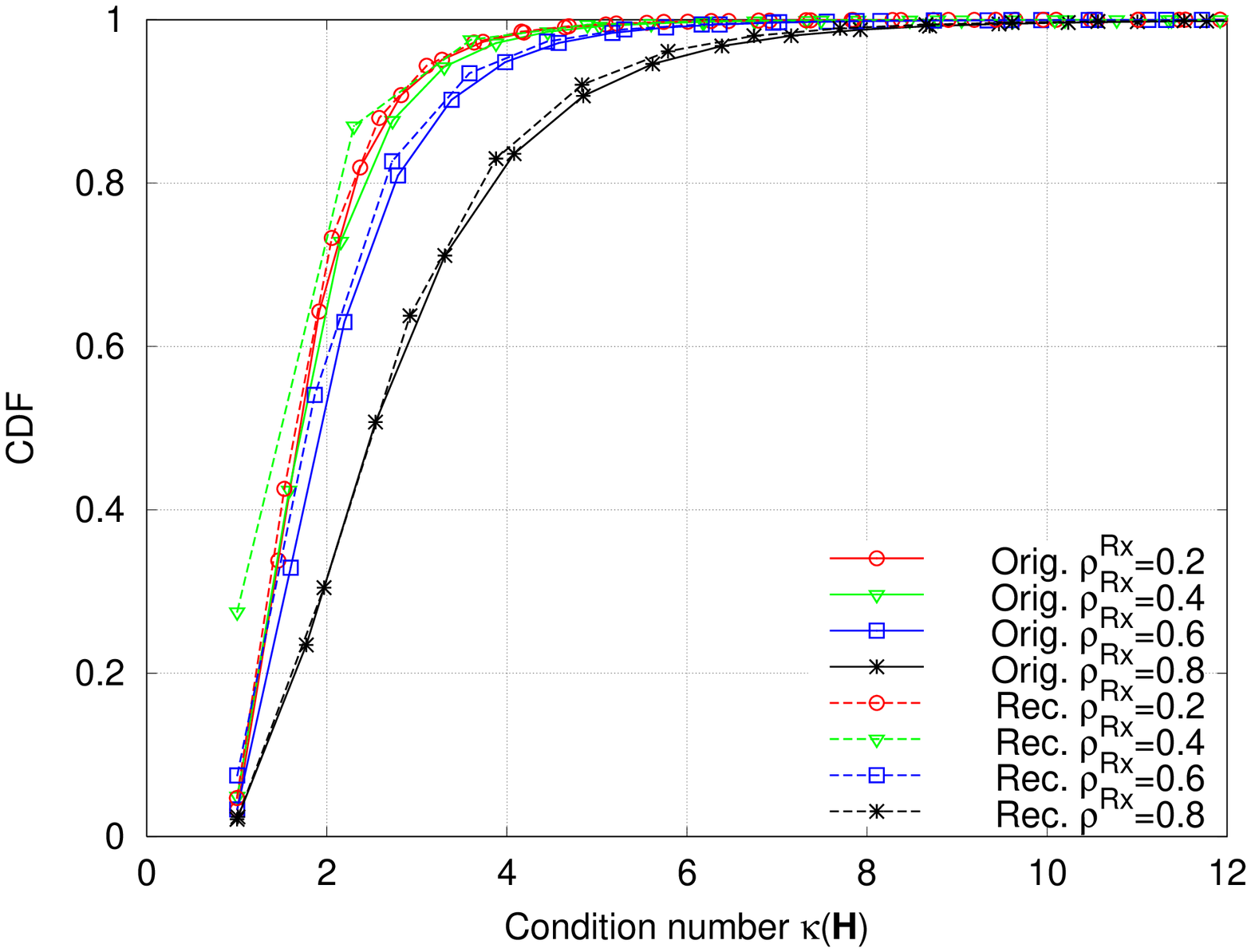}}
\subfigure[]{
   \includegraphics[width=0.45\textwidth]{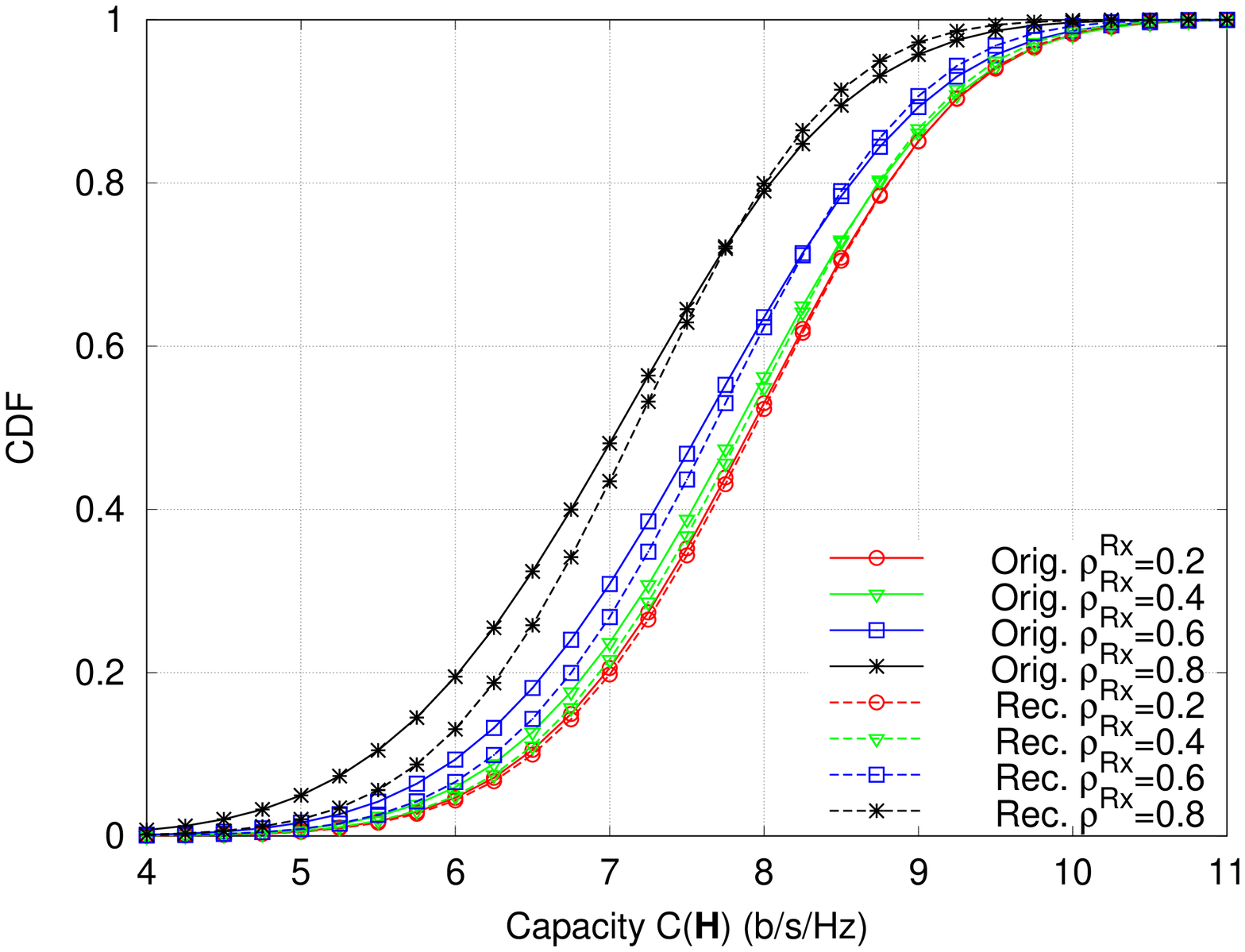}}
\caption{CDF distributions of time-independent scattering MIMO $2\times4$ channel for original and reconstructed channel matrices: (a) singular values, (b) channel condition numbers and (c) ergodic channel capacity calculated for SNR $\gamma = 20$ dB.}
\label{fig-cdf}
\end{figure*}

The comparison of cumulative distribution functions (CDF) of the original $\hat{\mathbf{H}}$ and reconstructed $\hat{\mathbf{H}}$ channel matrices is shown in Figure~\ref{fig-cdf}. Two distinct sets of singular values are present in Figure~\ref{fig-cdf}~(a) indicating channel of rank two. The close resemblance exists between statistics of eigenvalues (especially the smaller ones), channel condition numbers and ergodic capacity. Noticeable deviation between original and reconstructed channel statistics appears only for Rx correlation coefficient $\rho^{\rx} = 0.8$ which represent highly correlated channel conditions less typical in MIMO applications.

\subsection{Static cluster scattering channel}

We will use scattering caused by static clusters in a spatial (angular) domain within MIMO system which will allow us to retain correlation properties of reconstructed covariance matrix. The same cluster geometry will induce the same channel scattering behavior both for full-dimensional MIMO system, as well as for partial (reduced dimensionality) MIMO subsystems. This will result in similar angular scattering characteristics but with reduced spatial resolution for partial MIMO subsystem. We will analyze similarity of covariance matrices as well as power angular spectra between MIMO systems with full and reduced number of antennas.

For cluster scattering model we reconstruct higher order channel matrices from lower order correlation matrices and compare angle of arrival (AoA) and angle of departure (AoD) spectra. Consider equidistant linear antenna arrays at Rx and Tx with antenna element separations $d_{\rx}$ and $d_{\tx}$, respectively. Scattering occurs between Tx and Rx due to a number $K$ of resolvable clusters. Full channel correlation matrix can be represented by eigenvectors $\mathbf{u}_{k}$, $k = 1, \ldots, K$ \cite{WeichselbergerThesis03}
\begin{align}
  \mathbf{R}_{\mathrm{H}} &= \sum_{k=1}^{K} \mathbf{u}_{k} \mathbf{u}_{k}^{\tsup{T}}, \label{eq-cor-h-eigv} \\
  \mathbf{u}_{k} &= \rvec\left(\mathbf{U}_{k}\right), \\
  \mathbf{U}_{k} &= \mathbf{a}_{\rx}\left(\varphi_{\mathrm{AoA} \, k}\right) \mathbf{a}_{\tx}\left(\varphi_{\mathrm{AoD} \, k}\right)^{\tsup{T}}.
\end{align}
Here $\mathbf{a}_{\rx}$ and $\mathbf{a}_{\tx}$ are antenna array steering vectors at Rx and Tx sites with respect to cluster scattering angles $\varphi_{\mathrm{AoA} \, k}$ and $\varphi_{\mathrm{AoD} \, k}$:
\begin{align}
  \mathbf{a}_{\rx}\left(\varphi_{\mathrm{AoA} \, k}\right) &= \left[ \begin{array}{c}
  1 \\
  \exp\left[2\pi j\, \frac{d_{\rx}}{\lambda} \sin\varphi_{\mathrm{AoA} \, k} \right]  \\
  \exp\left[2\pi j \, 2 \, \frac{d_{\rx}}{\lambda} \sin\varphi_{\mathrm{AoA} \, k} \right]  \\
  \vdots \\
  \exp\left[2\pi j (N_{\rx} - 1) \frac{d_{\rx}}{\lambda} \sin\varphi_{\mathrm{AoA} \, k} \right]  \\
  \end{array} \right], \label{eq-a-rx}\\
  \mathbf{a}_{\tx}\left(\varphi_{\mathrm{AoD} \, k}\right) &= \left[ \begin{array}{c}
  1 \\
  \exp\left[2\pi j \, \frac{d_{\tx}}{\lambda} \sin\varphi_{\mathrm{AoD} \, k} \right]  \\
  \exp\left[2\pi j \, 2 \, \frac{d_{\tx}}{\lambda} \sin\varphi_{\mathrm{AoD} \, k} \right]  \\
  \vdots \\
  \exp\left[2\pi j (N_{\tx} - 1) \, \frac{d_{\tx}}{\lambda} \sin\varphi_{\mathrm{AoD} \, k} \right]  \\
  \end{array} \right], \label{eq-a-tx}
\end{align}
which are used to generated random channel matrix $\mathbf{H}$ using Cholesky factorization (\ref{eq-chol-fact}). In this way, random Rayleigh-distributed channel realizations $\mathbf{H}$ can be generated for MIMO $2\times2$ and $2\times4$ channels.

First consider having MIMO channel configuration $N_{\tx} \times N_{0}$ comprised of $N_{\tx}$ transmit and $N_{0}$ receive antennas as a baseline configuration and the goal is to estimate performance of an extended MIMO configuration $N_{\tx} \times N_{1}$ with the number of receivers $N_{1}$, such that $N_{1} > N_{0}$. Assuming Kronecker separability, the baseline MIMO channel spatial covariance matrix can be split into Rx and Tx parts as
\begin{align}
   \mathbf{R}_{\rx} &= \frac{1}{N_{\tx}} \mathbb{E}\left\{ \mathbf{H}^{\tsup{H}} \mathbf{H} \right\}, \\
   \mathbf{R}_{\tx} &= \frac{1}{N_{0}} \mathbb{E}\left\{ \mathbf{H} \mathbf{H}^{\tsup{H}} \right\}.
\end{align}
Suppose that channel measurements are available for the baseline MIMO configuration and its correlation matrices can be deduced from measurements. The rest of $N_{1}-N_{0}$ receive antennas can be estimated according to the scattering properties of clusters on linear antenna arrays expressed by steering vectors (\ref{eq-a-rx}) and (\ref{eq-a-tx}). Having capabilities of measuring $2 \times 2$ MIMO configurations with Rx antenna separations $d_{\rx} \cdot N_{0}, \ldots, d_{\rx} \cdot (N_{1}-1)$ and corresponding partial Rx correlation matrices $\mathbf{R}_{\rx \, 2 \times 2}(d_{\rx} \cdot n)$, $n = N_{0}, \ldots, N_{1}-1$, extended Rx correlation matrix can be reconstructed as
\begin{equation}
\hat{\mathbf{R}}_{\rx} = \left[\begin{array}{cccc|cccc}
		& & & & \rho_{1 \, (N_{0}+1)} & \rho_{1 \, (N_{0}+2)} & \ldots & \rho_{1 \, N_{1}}\\
        & & & & \rho_{2 \, (N_{0}+1)} & \rho_{2 \, (N_{0}+2)} & \ldots & \rho_{2 \, N_{1}}\\
        & & & & \vdots & \vdots & \ddots & \vdots \\
	\multicolumn{4}{c|}{\smash{\raisebox{1.5\normalbaselineskip}{$\left[\mathbf{R}_{\rx}\right]_{N_{0} \times N_{0}}$}}}
         & \rho_{N_{0} \, (N_{0}+1)} & \rho_{N_{0} \, (N_{0}+2)} & \ldots & \rho_{N_{0} \, N_{1}}\\
    \hline \\[-\normalbaselineskip]
		   \rho_{(N_{0}+1) \, 1} & \rho_{(N_{0}+1) \, 2} & \ldots & \rho_{(N_{0}+1) \, N_{0}} & 1 & \rho_{1 \, 1} & \ldots & \rho_{1 \, (N_{1}-N_{0})}\\
		   \rho_{(N_{0}+2) \, 1} & \rho_{(N_{0}+2) \, 2} & \ldots & \rho_{(N_{0}+2) \, N_{0}} & \rho_{2 \, 1} & 1 & \ldots & \rho_{2 \, (N_{1}-N_{0})}\\
		   \vdots & \vdots & \ddots & \vdots  & \vdots & \vdots & \ddots & \vdots \\
		   \rho_{N_{1} \, 1} & \rho_{N_{1} \, 2} & \ldots & \rho_{N_{1} \, N_{0}} & \rho_{N_{1} \, 1} & \rho_{N_{1} \, 2} & \ldots & 1\\
    \end{array}\right]_{N1 \times N1}, \label{eq-corr-rec-stat}
\end{equation}
where
\begin{equation}
	\rho_{mn} = [\mathbf{R}_{\rx}(d_{\rx}\cdot(m-n))]_{12} \label{eq-rho-dmn}
\end{equation}
are off-diagonal elements of partial MIMO $2\times 2$ receive correlation matrices $\mathbf{R}_{\rx}$ for variable antenna spacing $d_{\rx}\cdot(m-n)$. Upper right and lower left blocks of (\ref{eq-corr-rec-stat}) are symmetric since $\rho_{mn} = \rho_{nm}^{*}$ and therefore both of them can be reconstructed from the same set of $\rho_{mn}$ coefficients (\ref{eq-rho-dmn}). The lower right block of (\ref{eq-corr-rec-stat}) has elements with subtracted $N_{0}$ index and since the correlation depends only on the difference $d_{\rx}\cdot(m-n)$ both antenna indices could be reduced to minimize the number of independent $\rho_{mn}$ coefficients. In case $N_{1} = 2N_{0}$, this lower right block coincides with the baseline correlation matrix block $\left[\mathbf{R}_{\rx}\right]_{N_{0} \times N_{0}}$ and reconstructed Rx covariance matrix simplifies to
\begin{equation}
\hat{\mathbf{R}}_{\rx} = \left[\begin{array}{cccc|cccc}
		& & & & \rho_{1 \, (N_{0}+1)} & \rho_{1 \, (N_{0}+2)} & \ldots & \rho_{1 \, N_{1}}\\
        & & & & \rho_{2 \, (N_{0}+1)} & \rho_{2 \, (N_{0}+2)} & \ldots & \rho_{2 \, N_{1}}\\
        & & & & \vdots & \vdots & \ddots & \vdots \\
	\multicolumn{4}{c|}{\smash{\raisebox{1.5\normalbaselineskip}{$\left[\mathbf{R}_{\rx}\right]_{N_{0} \times N_{0}}$}}}
         & \rho_{N_{0} \, (N_{0}+1)} & \rho_{N_{0} \, (N_{0}+2)} & \ldots & \rho_{N_{0} \, N_{1}}\\
    \hline \\[-\normalbaselineskip]
		   \rho_{(N_{0}+1) \, 1} & \rho_{(N_{0}+1) \, 2} & \ldots & \rho_{(N_{0}+1) \, N_{0}} &  &  &  & \\
		   \rho_{(N_{0}+2) \, 1} & \rho_{(N_{0}+2) \, 2} & \ldots & \rho_{(N_{0}+2) \, N_{0}} &  &  &  & \\
		   \vdots & \vdots & \ddots & \vdots  &  &  &  & \\
		   \rho_{N_{1} \, 1} & \rho_{N_{1} \, 2} & \ldots & \rho_{N_{1} \, N_{0}} & \multicolumn{4}{c}{\smash{\raisebox{1.5\normalbaselineskip}{$\left[\mathbf{R}_{\rx}\right]_{N_{0} \times N_{0}}$}}} \\
    \end{array}\right]_{N1 \times N1}. \label{eq-corr-rec-stat2}
\end{equation}

The reconstructed receiver correlation matrix combined with original Tx correlation matrix will give Kronecker-type full channel correlation matrix $\hat{\mathbf{R}}_{\mathrm{H}} = \hat{\mathbf{R}}_{\rx} \otimes \mathbf{R}_{\tx}$.

\begin{figure*}
\centering \subfigure[]{
   \includegraphics[width=0.4\textwidth]{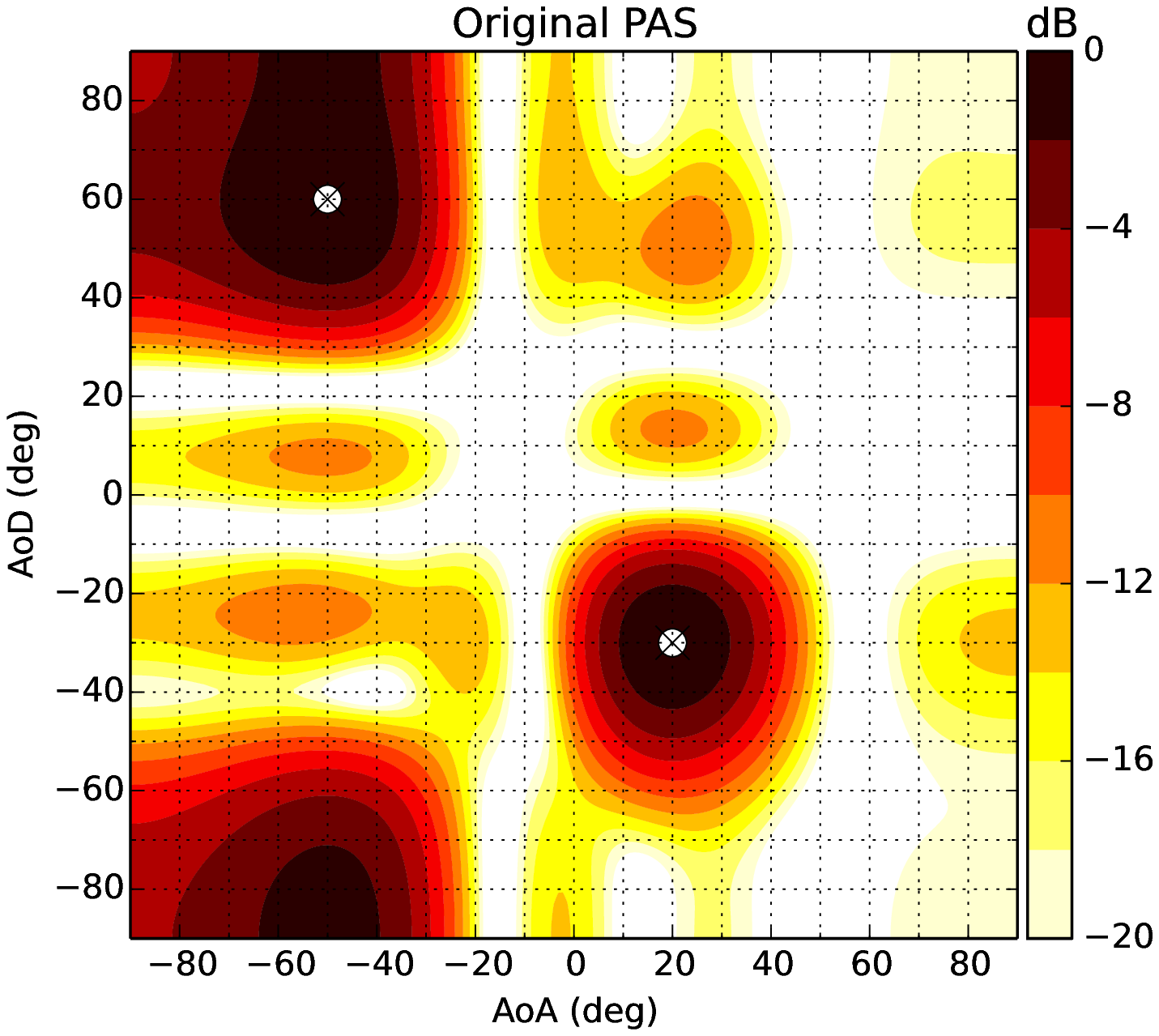}}
\subfigure[]{
   \includegraphics[width=0.4\textwidth]{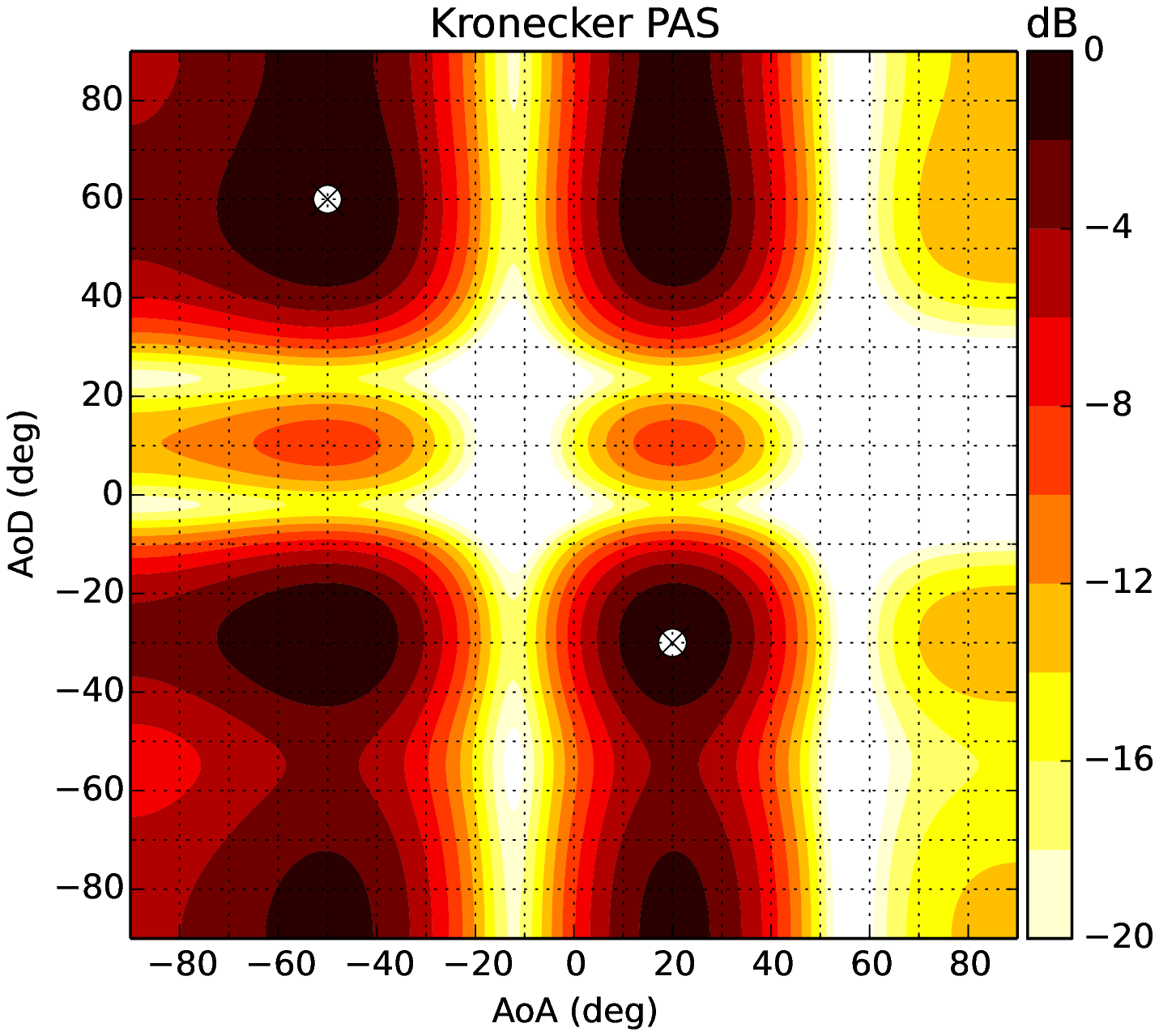}}
\subfigure[]{
   \includegraphics[width=0.4\textwidth]{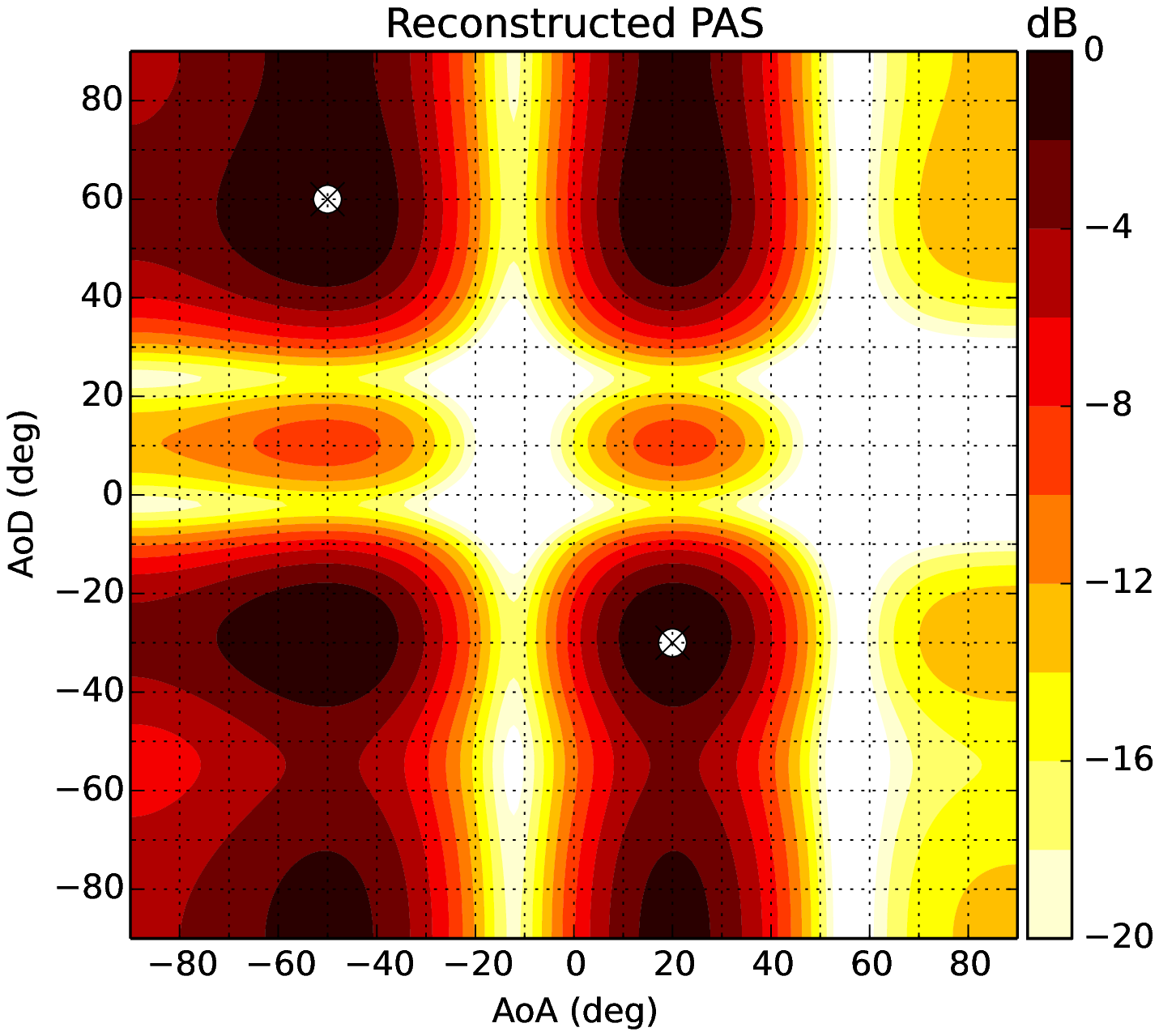}}
\subfigure[]{
   \includegraphics[width=0.4\textwidth]{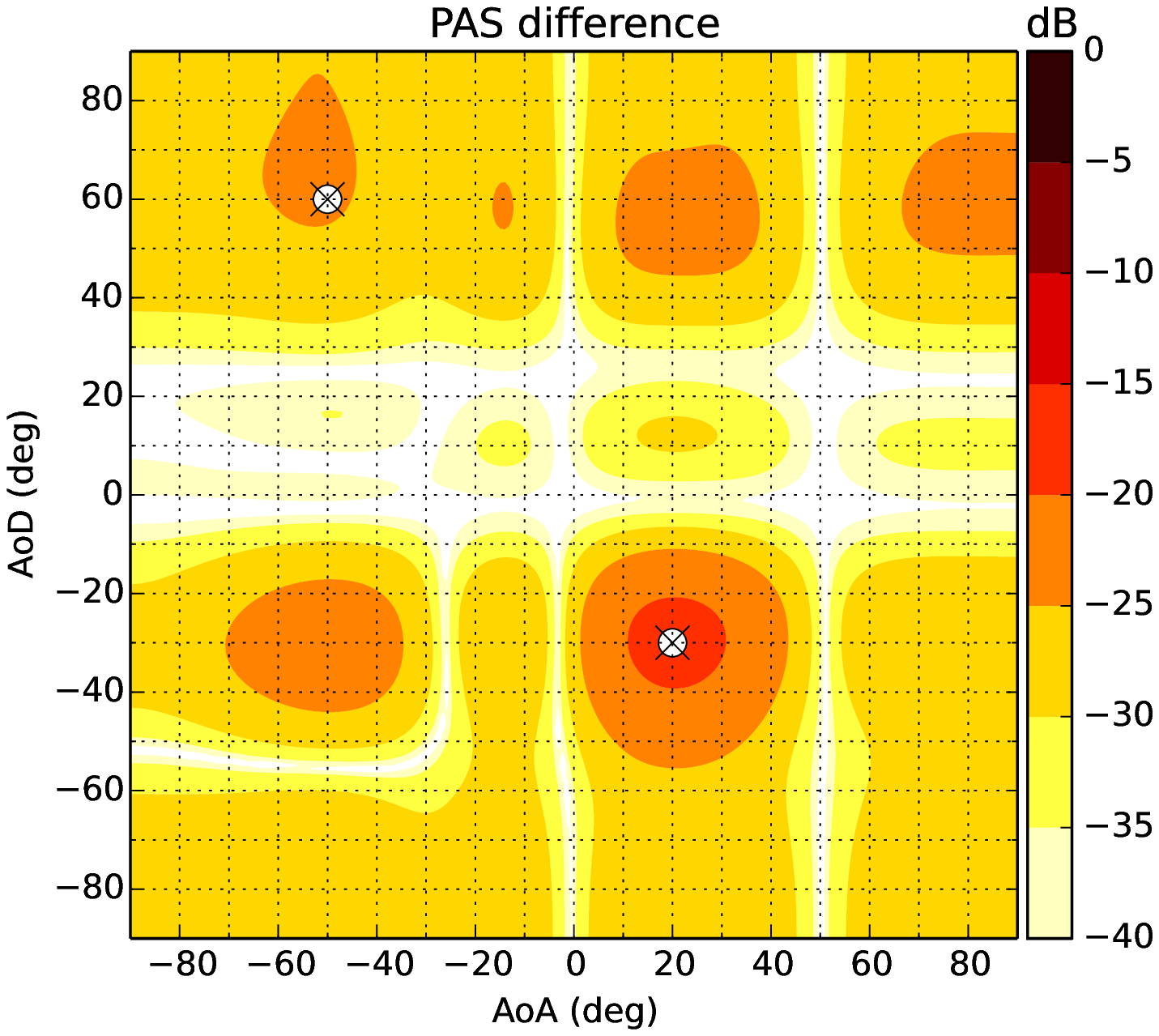}}
\caption{Power angular spectrum using Bartlett's beamformer for MIMO $N_{\tx} = 4$, $N_{\rx} = 8$, when antenna spacings are $d_{\rx} = 0.25 \lambda$, $d_{\tx} = 0.5 \lambda$: (a) original PAS, (b) PAS assuming Kronecker separation, (c) PAS reconstructed from $2\times2$ partial correlation matrices and (d) absolute difference between original PAS with Kronecker assumption and reconstructed PAS. Cross symbols ($\times$) denote true angular locations of scattering clusters.}
\label{fig-pas-24}
\end{figure*}

For a numerical example consider a baseline MIMO $4\times 4$ configuration subjected to  scattering from two clusters at $\left(\varphi_{\mathrm{AoA}}, \varphi_{\mathrm{AoD}} \right) = \left\{\left(-50^{\circ}, 60^{\circ} \right), \left(20^{\circ}, -30^{\circ} \right)\right\}$. The extended MIMO channel matrix have been constructed for MIMO $4 \times 8$ configuration with Gaussian fading generated according to spatial correlation matrix reconstructed from MIMO $4 \times 4$ channel and a series of partial $2 \times 2$ matrices for multiple Rx antenna separations. Since in this case $N_{1} = 2 \, N_{0}$, reconstructed correlation matrix can be obtained from (\ref{eq-corr-rec-stat2}). Reconstructed correlation matrix $\hat{\mathbf{R}}_{\mathrm{H}}$ has been compared to original MIMO $4 \times 8$ correlation matrix $\mathbf{R}_{\mathrm{H}}$ obtained from eigenvectors of the full channel correlation matrix (\ref{eq-cor-h-eigv}). By increasing the number of Gaussian random points in MIMO channel realization, reconstruction error can be minimized to desired level. For example MIMO channel snapshot with $10^{6}$ Gaussian points leads to the difference between the original and reconstructed Rx spatial correlation matrices with relative matrix error (\ref{eq-rel-err}), CMD (\ref{eq-cmd}) and CMC (\ref{eq-cmc}) having the following values: $\epsilon\left(\hat{\mathbf{R}}_{\rx}, \mathbf{R}_{\rx}\right) = 0.9\cdot10^{-3}$, $d_{\mathrm{corr}}\left(\hat{\mathbf{R}}_{\rx}, \mathbf{R}_{\rx}\right) = 4\cdot 10^{-7}$ and $c\left(\hat{\mathbf{R}}_{\rx}, \mathbf{R}_{\rx}\right) = 1 - 1.2\cdot 10^{-3}$.

At the same time close similarity exists between original and reconstructed PAS spectra. Fig.~\ref{fig-pas-24} shows original, Kronecker separated and reconstructed PAS for MIMO $4 \times 8$ configuration obtained using $10^{6}$ Gaussian points. The angular spectra have been generated using Bartlett's beamformer. The difference between original Kronecker based PAS and reconstructed PAS is shown in Fig.~\ref{fig-pas-24}~(d) with the maximal difference of -18.9~dB and the mean difference being -27.4~dB. This indicates possibility of recovering PAS characteristics using lower order partial MIMO subsystems until spatial resolution enables discerning the number of existing clusters.

\subsection{Time-variant cluster scattering channel}

\begin{figure}
  \centering
  \includegraphics[width=0.6\textwidth]{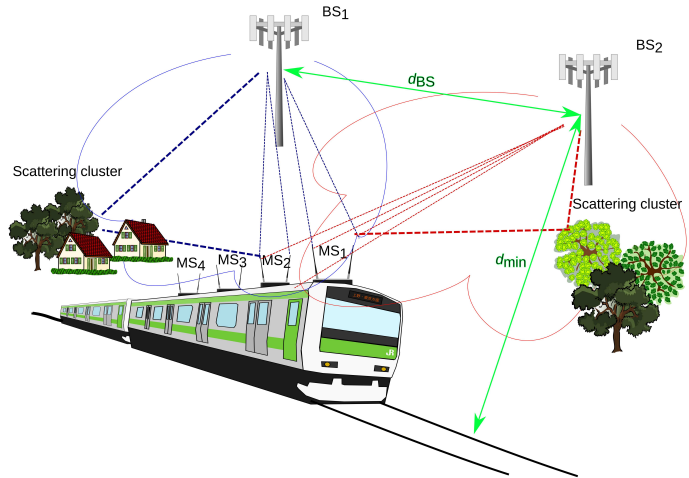}
\caption{Geometry of high-speed train scattering channel scenario.}
\label{fig-hst}
\end{figure}

As an example of time-variant scattering channel we use simulated high-speed train (HST) scenario shown in Fig.~\ref{fig-hst}. There is some degree of correlation between power angular spectra and power delays of adjacent mobile station (MS) antennas mounted on the train assuming there exist same scattering clusters for adjacent train antennas, similar cluster visibility regions and Doppler spreads.

We build a discrete stochastic channel model following methods of \cite{KunnariMIMOModel07}, \cite{XiaoKron304}, \cite{GhazalMIMONonstationary15} limiting analysis to narrowband Rician channel, although wideband case is readily extensible based on \cite{KunnariMIMOModel07}, \cite{GhazalMIMONonstationary15}. Rician channel model is constructed combining line-of-sight (LOS) part $\mathbf{H}_{\mathrm{LOS}}(t, f)$ and non-line-of-sight (NLOS) part $\mathbf{H}_{\mathrm{NLOS}}(t, f)$ as a $N_{\tx}\times N_{\rx}$ matrix functions over time and frequency:
\begin{align}
  &\mathbf{H}(t, f) = \sqrt{\frac{K_{\mathrm{Rician}}}{K_{\mathrm{Rician}}+1}} \mathbf{H}_{\mathrm{LOS}}(t, f) + \sqrt{\frac{1}{K_{\mathrm{Rician}}+1}} \mathbf{H}_{\mathrm{NLOS}}(t, f), \label{eq-hmx-rician}\\
  &\mathbf{H}_{\mathrm{LOS}}(t, f) = \mathbf{a}_{\tx}^{\tsup{T}}\left(\varphi_{\mathrm{AoD}}\right) \otimes \mathbf{a}_{\rx}\left(\varphi_{\mathrm{AoA}}\right) \exp\left\{j2\pi \frac{v}{\lambda} \cos\theta_{v} t - f\tau_{0}\right\}, \\
  &\rvec\left(\mathbf{H}_{\mathrm{NLOS}}(t, f)\right) = \mathbf{R}_{\mathrm{H}}^{1/2} \rvec\left(\mathbf{H}_{\mathrm{iid}}\right),
\end{align}
where $K_{\mathrm{Rician}}$ is Rician factor, $v = |\mathbf{v}|$ is the velocity, $\theta_{v}$ is the angle between LOS path and velocity vector $\mathbf{v}$, $\tau_{0}$ is the path delay of LOS component, $\lambda$ is the wavelength, while steering vectors $\mathbf{a}_{\rx}$ and $\mathbf{a}_{\tx}$ are given by (\ref{eq-a-rx}) and (\ref{eq-a-tx}) for LOS path visibility angles $\varphi_{\mathrm{AoA}}$ and $\varphi_{\mathrm{AoD}}$.

For estimating NLOS part $\mathbf{H}_{\mathrm{NLOS}}(t, f)$ of the channel, full channel correlation matrix approximation using Kronecker decomposition
\begin{equation}
  \mathbf{R}_{\mathrm{H}} = \mathbf{R}_{\rx} \otimes \mathbf{R}_{\tx}
\end{equation}
is obtained from Rx and Tx correlation matrices
\begin{equation}
  \mathbf{R}_{\rx} = \left[\rho\left(d_{\rx}\cdot(m-n)\right) \right]_{N_{\rx}\times N_{\rx}}, \quad
  \mathbf{R}_{\tx} = \left[\rho\left(d_{\tx}\cdot(m-n)\right) \right]_{N_{\tx}\times N_{\tx}}, \label{eq-rx-tx-corr-mx}
\end{equation}
where $d_{\rx}$ and $d_{\tx}$ are, respectively, antenna element separations for linear Rx and Tx antenna arrays. Here we assume continuous distribution within cluster based on Gaussian angular distribution over AoA or AoD angle $\varphi$ as
\begin{align}
  p(\varphi) = \begin{cases}
      \frac{1}{\sqrt{2\pi}\sigma_{\phi}} \exp\left\{-\frac{\left(\varphi - \bar{\varphi}\right)^{2}}{2\sigma_{\varphi}^{2}} \right\}, & \text{if $\left|\varphi - \bar{\varphi}\right|$} \le \pi \\
      0, & \text{otherwise}
    \end{cases}
\end{align}
where $\bar{\varphi}$ is the mean and $\sigma_{\phi}$ is the standard deviation of AoA or AoD angle. Following \cite{Buehrer02}, correlation coefficients between adjacent MIMO antennas can be derived analytically as functions over the distance $d_{mn}$ between antenna elements $m$ and $n$:
\begin{align}
  \rho(d_{mn}) &= \int_{\bar{\varphi}-\pi}^{\bar{\varphi}+\pi}{\exp\left\{j 2\pi \frac{d_{mn}}{\lambda} \sin\varphi \right\} p(\varphi) \rmd \varphi} \nonumber \\
  &= \exp\left\{ j 2\pi \frac{d_{mn}}{\lambda} \sin\bar{\varphi} \right\} \exp\left\{ -\frac{1}{2}\left( 2\pi\frac{d_{mn}}{\lambda}\sigma_{\varphi} \cos\varphi \right)^{2} \right\}, \label{eq-rho-dmn-hst}
\end{align}
which are elements of Rx and Tx correlation matrices (\ref{eq-rx-tx-corr-mx}).

For practical purposes time-varying frequency response can be estimated from sampled version of channel function \cite{Bernado14} over time and frequency domains represented, respectively, by indexes $i$ and $k$:
\begin{equation}
  H_{ik} \equiv H[i, k] = H(i T_{\mathrm{s}}, k F_{\mathrm{s}}), \label{eq-H-sampled}
\end{equation}
where $T_{\mathrm{s}}$ and $F_{\mathrm{s}}$ are time and frequency domain sampling steps. Here spatial dimensions $m$, $n$ of channel matrix are omitted for simplicity reasons and the following time-variant analysis applies to each transmit-receive matrix element. We use local scattering function (LSF) $\pazocal{C}_{\mathrm{H}}(t, f; \tau, \nu)$ over time $t$, frequency $f$, delay $\tau$ and Doppler frequency $\nu$ to describe correlation properties of time-varying channel over short stationarity periods when wide-sense stationary uncorrelated scattering (WSSUS) assumption is satisfied. LSF for nonstationary channel can be constructed following \cite{MatzNonWSSUS05}, \cite{PaierWSSUS08} as an integral over time and delay lags, $\Delta t$ and $\Delta \tau$,
\begin{equation}
  \pazocal{C}_{\mathrm{H}}(t, f; \tau, \nu) = \int_{-\infty}^{\infty}\int_{-\infty}^{\infty} {R_{\mathrm{h}}(t, \tau; \Delta t, \Delta \tau) \rme^{-j 2\pi (\nu\Delta t + f \Delta\tau)} \rmd\Delta t \rmd\Delta \tau},
\end{equation}
of four-dimensional correlation function of the impulse response $h(t, \tau)$:
\begin{equation}
  R_{\mathrm{h}}(t, \tau; \Delta t, \Delta \tau) = \mathbb{E}\left\{h(t, \tau+\Delta\tau) h^{*}(t-\Delta t, \tau) \right\}.
\end{equation}

For numerical analysis a generalized version of LSF \cite{MatzNonWSSUS05}, \cite{PaierWSSUS08} is used which is a smooth and localized function about the origin in time-frequency plane consisting of $N_{w}$ linearly independent prototype systems $G_{w}$:
\begin{align}
  \pazocal{C}_{\mathrm{H}}^{(\Phi)}(t, f; \tau, \nu) &= \mathbb{E}\left\{\sum_{w=0}^{N_{w}-1}{\gamma_{w} \left| \pazocal{H}^{(G_{w})}(t, f; \tau, \nu) \right|^{2} } \right\}, \\
  \pazocal{H}^{(G_{w})}(t, f; \tau, \nu) &= \rme^{j 2\pi f \tau} \int_{-\infty}^{\infty} \int_{-\infty}^{\infty} {L_{\mathrm{H}} (t', f') L_{G_{w}}^{*}(t'-t, f'-f) \rme^{-j 2\pi \left(\nu t' - \tau f' \right)} \rmd t' \rmd f'},
\end{align}
where $L_{\mathrm{H}} (t', f')$ is time-frequency channel transfer function \cite{MatzNonWSSUS05} and $L_{G_{w}}^{*}(t'-t, f'-f)$ is transfer function of temporary localized low-pass filters $G_{w}$, which can be implemented numerically by discrete time-frequency function
\begin{equation}
  L_{G_{w}}[i, k] = u_{i}\left[i + M_{t}/2\right] \tilde{u}_{k}\left[k + M_{f}/2\right]
\end{equation}
of the discrete prolate spheroidal sequences $u_{i}[i']$ and $\tilde{u}_{k}[k']$ \cite{MatzNonWSSUS03}, \cite{Slepian78} concentrated in time-frequency region $i' = 0, \ldots, M_{t}-1$ and $k' = 0, \ldots, M_{f}-1$. Here $M_{t}$ and $M_{f}$ denote number of tapers in time and frequency domains, respectively. Coefficients $\gamma_{w}$ are arbitrarily chosen up to condition $\sum_{w=0}^{N_{w}-1}{\gamma_{w}} = 1$. Discrete version of windowed multi-taper correlation function constructed using number $I_{w}$ of time-domain and number $K_{w}$ of frequency-domain orthogonal tapers:
\begin{align}
  \pazocal{C}[m_{t}, m_{f}; l, p] &= \frac{1}{I_{w} K_{w}} \sum_{w=0}^{I_{w} K_{w}} {\left| \pazocal{H}_{G_{w}}[m_{t}, m_{f}; l, p] \right|^{2}}, \label{eq-sampled-corr}\\
  \pazocal{H}_{G_{w}}[m_{t}, m_{f}; l, p] &= \sum_{i'=-M_{t}/2}^{M_{t}/2-1}\sum_{k'=-M_{f}/2}^{M_{f}/2-1} {H[i'-m_{t}, k'-m_{f}] G_{w}[i', k'] \rme^{-j 2\pi (pi'-lk')}}, \\
  G_{w}[i', k'] &= u_{i}\left[i' + M_{t}/2\right] \tilde{u}_{k}\left[k' + M_{f}/2\right],
\end{align}
where composite two-dimensional window index $w = i K_{w} + k$, $i = 0, \ldots, I_{w}-1$, $k = 0, \ldots, K_{w}-1$ and $H[i, k]$ is the sampled channel matrix (\ref{eq-H-sampled}).

The total number of time snapshots $N_{\mathrm{ts}}$ and frequency subcarriers $N_{\mathrm{sc}}$ are divided into $M_{t} \times M_{f}$ size stationarity regions each of which is identified by time index $m_{t} = 1, \allowbreak \ldots, \allowbreak N_{\mathrm{ts}}/M_{t}-1$ and frequency index $m_{f} = 1, \ldots, N_{\mathrm{sc}}/M_{f}-1$. Relative indexes within stationarity regions are defined as:
\begin{align}
  i' &= -M_{t}/2, \ldots, M_{t}/2-1, \quad i = m_{t} M_{t} + i', \\
  k' &= -M_{f}/2, \ldots, M_{f}/2-1, \quad k = m_{f} M_{f} + k'.
\end{align}
Delay index is limited to region $l = 0, \ldots, M_{f}-1$ and Doppler index varies over $p = -M_{t}/2, \allowbreak \ldots, \allowbreak M_{t}/2-1$.

\begin{figure}
  \centering
  \includegraphics[width=0.6\textwidth]{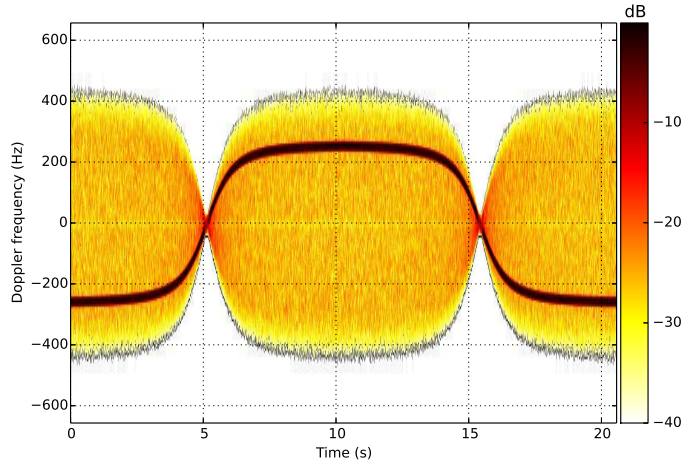}
\caption{Doppler power spectral density for HST scenario shown in Fig.~\ref{fig-hst}.}
\label{fig-hst-dsd}
\end{figure}

The method of time-varying channel matrix reconstruction has been applied to HST scenario for train moving with constant velocity $v = 350$~km/h parallel to a row of base stations located at $d_{\min}=50$~m from railroad with inter-site distance of $d_{\mathrm{BS}}=1$~km (Fig.~\ref{fig-hst}). Frequency carrier is $f_{\mathrm{c}} = 800$~MHz, Rician factor $K_{\mathrm{Rician}} = 0$~dB, Rx and Tx antenna array separations are $d_{\rx} = d_{\tx} = \lambda/2$, AoA and AoD angle deviation in NLOS scattering is set to $\sigma_{\varphi} = 20^{\circ}$. In order to reflect time-dependent channel behavior, Doppler power spectral density (DSD) is calculated as an average of sampled correlation function (\ref{eq-sampled-corr}) over delays \cite{Bernado14}:
\begin{equation}
DSD[m_{t}, m_{f}; p] = \frac{1}{N_{f}} \sum_{l=0}^{M_{f}-1} {\pazocal{C}[m_{t}, m_{f}; l, p]}. \label{eq-dsd}
\end{equation}
For the frequency variable $m_{f}$ fixed to a central subcarrier, DSD is a two-dimensional function over time and Doppler domains indexed by $m_{t}$ and $p$, respectively.

Due to varying AoA angle for train moving along the railroad, Doppler frequency varies with time as \cite{3GPP-TS36.104-16}
\begin{align}
  &\nu_{\mathrm{D}}(t) = f_{\mathrm{c}} \frac{v}{c} \cos\varphi_{\mathrm{AoA}}(t), \label{eq-hst-v-doppler}\\
  &\cos\varphi_{\mathrm{AoA}}(t) = \begin{cases}
      \frac{d_{\mathrm{BS}}/2 - vt}{\sqrt{d_{\min}^{2} + \left(d_{\mathrm{BS}}/2 - vt \right)^{2}}}, & 0 \le t \le d_{\mathrm{BS}}/v\\
      \frac{-3 d_{\mathrm{BS}}/2 + vt}{\sqrt{d_{\min}^{2} + \left(-3 d_{\mathrm{BS}}/2 + vt \right)^{2}}}, & d_{\mathrm{BS}}/v < t \le 2 d_{\mathrm{BS}}/v \\
      \cos \varphi_{\mathrm{AoA}}\left(t \bmod\left(2 d_{\mathrm{BS}}/v\right) \right),  & t > 2 d_{\mathrm{BS}}/v,
    \end{cases} \label{eq-hst-v-cos}
\end{align}
where $c$ is the speed of light.

This time-dependent Doppler frequency behavior is reflected in Fig.~\ref{fig-hst-dsd} displaying DSD calculated using random Rician channel realizations according to (\ref{eq-dsd}). Time domain has been sampled by $N_{\mathrm{ts}} = 2^{21}$ points and split into 2048 stationarity regions each covered by $M_{t} = 1024$ tapers. The meander-like dark line corresponds to LOS part of the channel which closely follows Doppler frequency temporal dependence due to varying AoA angle (\ref{eq-hst-v-cos}). The spread of DSD both in positive and negative Doppler frequencies is characteristic to classical Doppler spectrum used for generating NLOS channel part.

\begin{figure*}
  \centering
  \subfigure[]{
   \includegraphics[width=0.5\textwidth]{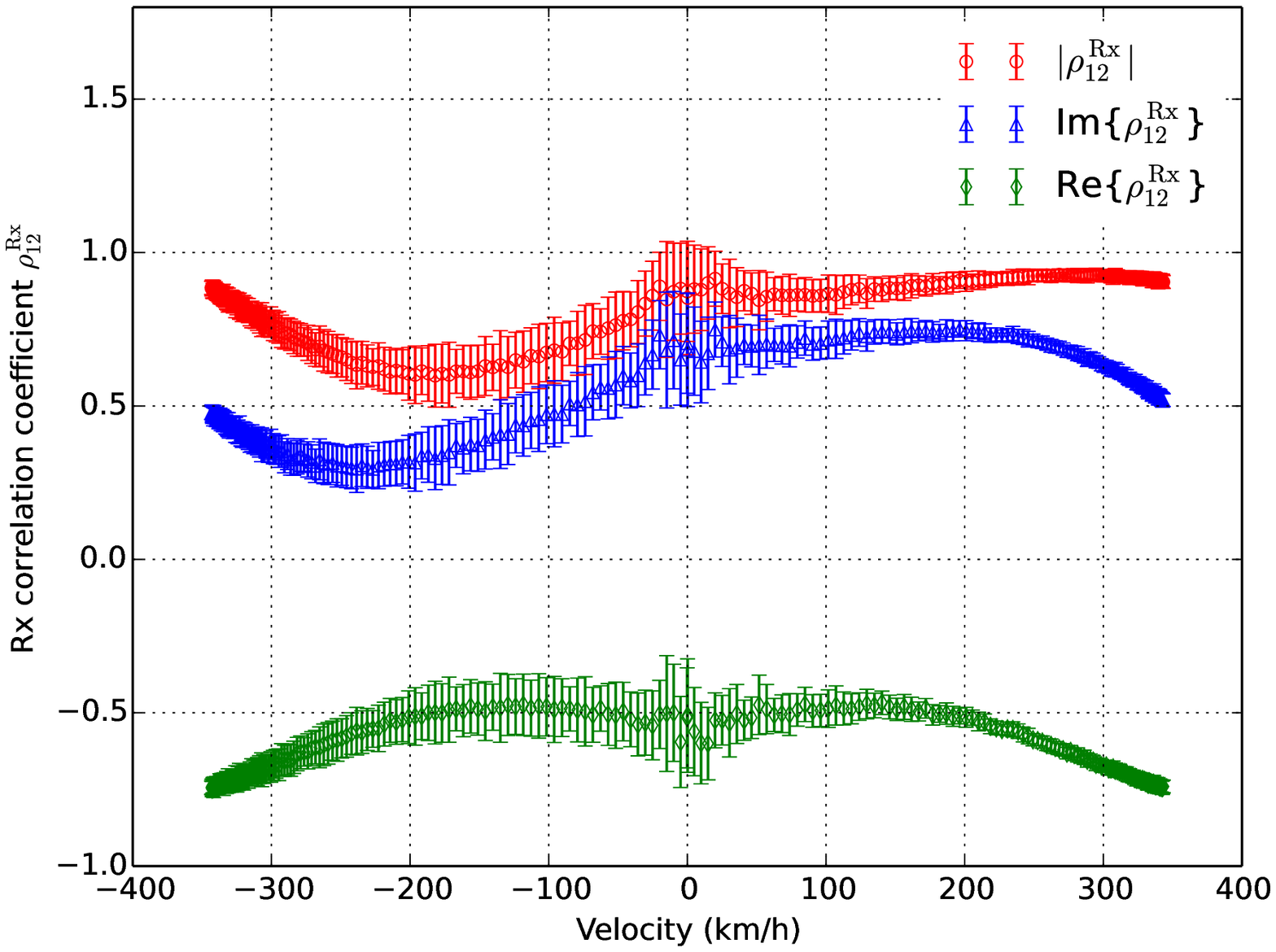}}
  \subfigure[]{
   \includegraphics[width=0.47\textwidth]{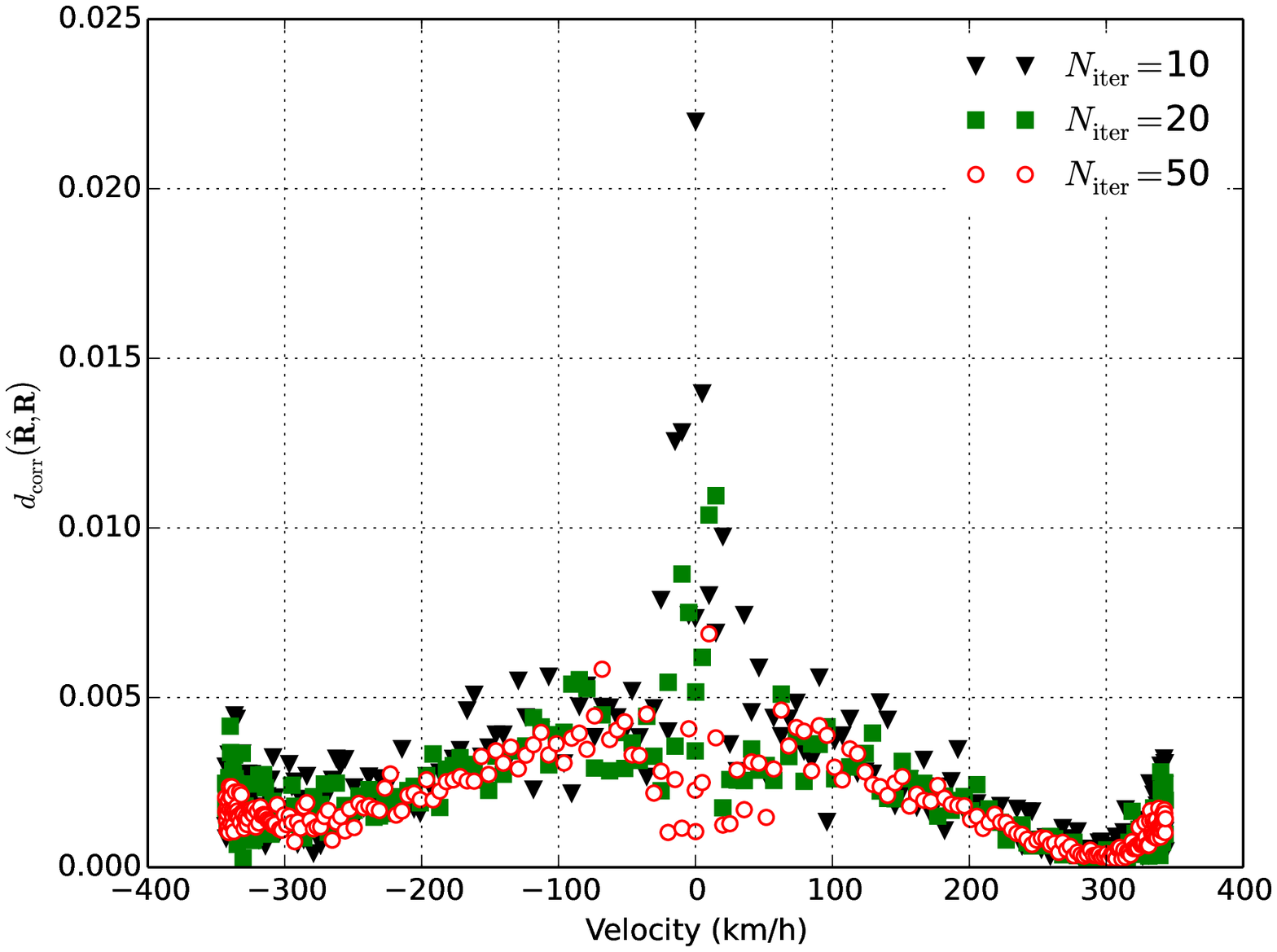}}
\caption{Complex correlation coefficient $\rho_{12}^{\mathrm{\rx}}$ between two adjacent receiving antenna elements for each stationarity region in HST scenario (a). Error bars correspond to $1.96 \sigma$ variation or 0.95 level of confidence interval of the correlation coefficient variation over randomized channel snapshots generated within $N_\mathrm{iter} = 50$ iterations. Channel matrix reconstruction CMD error $d_{\mathrm{corr}}\left(\hat{\mathbf{R}}, \mathbf{R}\right)$ for the same scenario but different number of iterations is shown in (b).}
\label{fig-time-var-rho}
\end{figure*}

Reconstructed Rx correlation matrix $\hat{\mathbf{R}}_{\rx}$ can be obtained by applying decomposition (\ref{eq-corr-rec-stat2}) in each stationarity region where WSSUS assumption holds. In order to build $4\times 4$ sized Rx correlation matrix $\hat{\mathbf{R}}_{\rx}$ for MIMO $2\times 4$ configuration, we use a combination of three simulated MIMO $2\times2$ channel correlation matrices generated with fixed Tx antenna separation $d_{\tx}$ and multiple Rx antenna separations, $d_{\rx}$, $2 \cdot d_{\rx}$ and $3 \cdot d_{\rx}$, resulting in correlation coefficients $\rho_{12}^{\rx}$, $\rho_{13}^{\rx}$ and $\rho_{14}^{\rx}$, respectively, with the use of (\ref{eq-rx-tx-corr-mx}) and (\ref{eq-rho-dmn-hst}). In case of four receivers, reconstructed correlation matrix takes the form of
\begin{equation}
\hat{\mathbf{R}}_{\rx} = \left[ \begin{array}{cccc}
 1       &  \rho^{\rx}_{12} & \rho^{\rx}_{13} & \rho^{\rx}_{14} \\
 {\rho^{\rx}_{12}}^{*} &  1   &  \rho^{\rx}_{12} & \rho^{\rx}_{13} \\
{\rho^{\rx}_{13}}^{*} &  {\rho^{\rx}_{12}}^{*} &  1  & \rho^{\rx}_{12} \\
{\rho^{\rx}_{14}}^{*} & {\rho^{\rx}_{13}}^{*} &  {\rho^{\rx}_{12}}^{*} & 1
\end{array} \right].
\end{equation}
Within validity of Kronecker decomposition, reconstructed $4\times 4$ Rx correlation matrix is compared with the original correlation matrix $\mathbf{R}_{\rx}$ simulated using full- dimensional Rician channel (\ref{eq-hmx-rician}). The resulting complex correlation coefficient between two adjacent receive antennas $\rho_{12}^{\mathrm{\rx}}$ depends on relative velocity between Tx and Rx due to variability of Doppler shift over changing AoA angle. Variability of $\rho_{12}^{\mathrm{\rx}}$ over random channel realizations is shown in Fig.~\ref{fig-time-var-rho}~(a) over velocity range $\left[-v_{\mathrm{max}}, \ldots, v_{\mathrm{max}}\right]$, where maximum velocity $v_{\mathrm{max}} = 0.995 v$ corresponds to maximum Doppler shift in (\ref{eq-hst-v-doppler}) for a given HST geometry. Fig.~\ref{fig-time-var-rho}~(b) displays channel matrix reconstruction CMD error between original $\mathbf{R}_{\rx}$ and reconstructed $\hat{\mathbf{R}}_{\rx}$ correlation matrices. In this case, $N_{\mathrm{ts}} = 2^{18}$ time samples and $M_{t} = 1024$ time tapers have been used to reconstruct Rx correlation matrix. Although the reconstruction error depends on the train velocity as it moves along the railroad and reaches maximal values when relative velocity passes through zero value at the point where train is closest to the base station, maximal CMD error reduces when the number of iterations increases and falls below 0.01 for $N_\mathrm{iter} = 50$. This shows possibility of reconstructing MIMO channel correlation matrix from lower order partial correlation matrices in time-varying scenario with sufficient time domain sampling and channel statistics per stationarity region.

\section{Dual-polarized MIMO channel reconstruction}\label{sec-dual-pol}

Dual polarized MIMO antenna system represents multiplexing of spatial and polarimetric domains. There are numerous efforts to define statistical models of such channels \cite{Oestges08}, \cite{Liolis10}, \cite{Brown11}, \cite{EkpePhD12}, \cite{King12}. Channel transfer matrix for dual-polarization single antenna unit case can be defined as
\begin{equation}
\left[\mathbf{H}_{\times}\right]_{2\times2} = \left[ \begin{array}{cccc}
h_\mathsf{RR}  &  h_\mathsf{RL} \\
h_\mathsf{RL}  &  h_\mathsf{LL}
\end{array} \right],
\end{equation}
where matrix element indexes $\mathsf{R}$ and $\mathsf{L}$ denotes right and left hand circular polarized components.

We adopt decomposition of the full channel matrix into spatial and polarimetric components for Rayleigh distributed channel \cite{Oestges08}
\begin{equation}
\mathbf{H}_{\times} = \mathbf{H} \otimes \mathbf{X}, \label{eq-x-dec}
\end{equation}
where $\mathbf{H}$ is $N_{\rx}/2 \times N_{\tx}/2$ spatial channel matrix representing spatial multiplexing of MIMO streams, while $\mathbf{X}$ being $2\times2$ polarimetric channel matrix between the two dual-polarized antenna units.

In order to use partial channel measurements to reconstruct the full channel matrix, we split full channel correlation matrix
\begin{equation}
\mathbf{R}_{\mathrm{H}} = \mathbb{E}\left\{\rvec\left(\mathbf{H}_{\times}^{\tsup{H}}\right) \rvec\left(\mathbf{H}_{\times}^{\tsup{H}}\right)^{\tsup{H}}\right\}
\end{equation}
into spatial $\mathbf{R}_{\mathrm{S}}$ and polarimetric $\mathbf{R}_{\mathrm{P}}$ parts, the former involving averaging over polarizations and the later -- averaging over spatial locations of antennas:
\begin{align}
 \mathbf{R}_{\mathrm{S}} &= \mathbb{E}_{\substack{\text{Rx Pol} \in {\mathsf{R}, \mathsf{L}} \\ \text{Tx Pol} \in {\mathsf{R}, \mathsf{L}}}}\left\{\rvec\left(\left[\mathbf{H}_{\times}\right]_{N_{\rx}/2 \times N_{\tx}/2}^{\tsup{H}}\right) \rvec\left(\left[\mathbf{H}_{\times}\right]_{N_{\rx}/2 \times N_{\tx}/2}^{\tsup{H}}\right)^{\tsup{H}}\right\}, \label{eq-R-sp} \\
 \mathbf{R}_{\mathrm{P}} &= \mathbb{E}_{\substack{\text{Rx}=1, \ldots, N_{\rx} \\ \text{Tx}=1, \ldots, N_{\tx}}}\left\{\rvec\left(\left[\mathbf{H}_{\times}\right]_{2\times2}^{\tsup{H}}\right) \rvec\left(\left[\mathbf{H}_{\times}\right]_{2\times2}^{\tsup{H}}\right)^{\tsup{H}}\right\}. \label{eq-R-pol}
\end{align}
Here $\mathbf{H}_{\times}$ matrix dimensions inside vectorization functions are limited to single polarization for $\mathbf{R}_{\mathrm{S}}$ in (\ref{eq-R-sp}) allowing averaging over all polarizations, thus resulting in $N_{\rx}/2 \times N_{\tx}/2$ spatial dimensions. In case of $\mathbf{R}_{\mathrm{P}}$ correlation matrix (\ref{eq-R-pol}) -- averaging of polarimetric part of $\mathbf{H}_{\times}$ is performed over all spatial Rx and Tx locations.

We use Cholesky decomposition of spatial part of the channel matrix
\begin{equation}
\rvec(\mathbf{H}) = \mathbf{R}_{\mathrm{S}}^{1/2} \rvec\left(\mathbf{H}_{\mathrm{iid}}\right), \label{eq-chol-sp}
\end{equation}
where $\mathrm{H}_{\mathrm{iid}} \sim \pazocal{CN}(0,1)$ is circularly symmetric complex Gaussian random matrix. Following \cite{Oestges08} and \cite{ClerckxOestgesBook13}, polarimetric channel matrix can be approximated as
\begin{equation}
\rvec(\mathbf{X}) = \mathbf{R}_{\mathrm{P}}^{1/2} \rvec\left(\mathbf{X}_{w}\right), \label{eq-chol-pol}
\end{equation}
where $\mathbf{R}_{\mathrm{P}}$ is $4\times4$ correlation matrix between polarization states $\{\mathsf{R}, \mathsf{L}\}$ obtained by averaging over all transmit and receive antennas and $\mathbf{X}_{w}$ is a $2\times2$ matrix whose elements are independent
circularly symmetric complex exponentials:
\begin{equation}
  \mathbf{X}_{w} = \frac{1}{\sqrt{1 + \chi}} \left[ \begin{array}{cccc}
\rme^{j\phi_{\mathsf{R}\mathsf{R}}}  &  \sqrt{\chi\mu} \; \rme^{j\phi_{\mathsf{R}\mathsf{L}}} \\
\sqrt{\chi} \; \rme^{j\phi_{\mathsf{L}\mathsf{R}}}  &  \sqrt{\mu} \; \rme^{j\phi_{\mathsf{L}\mathsf{L}}}
\end{array} \right],
\end{equation}
where $\mu$ and $\chi$ correspond to the inverse of co-polar (CPR) and cross-polar (XPR) ratios:
\begin{equation}
\mu^{-1} = \mathrm{CPR} = \frac{\mathbb{E}\left\{\left|h_{\mathsf{R}\mathsf{R}}\right|^{2}\right\}}{\mathbb{E}\left\{\left|h_{\mathsf{L}\mathsf{L}}\right|^{2}\right\}}, \quad
\chi^{-1} = \mathrm{XPR} = \frac{\mathbb{E}\left\{\left|h_{\mathsf{R}\mathsf{R}}\right|^{2}\right\}}{\mathbb{E}\left\{\left|h_{\mathsf{R}\mathsf{L}}\right|^{2}\right\}}.
\end{equation}
Phases of polarization components $\phi_{p,q}$, $p,q \in \{\mathsf{R}, \mathsf{L}\}$ are randomly distributed over $[0, 2\pi)$ interval and $\mu$ and $\chi$ are distributed as lognormal variables \cite{ClerckxOestgesBook13}. After generating spatial (\ref{eq-chol-sp}) and polarimetric (\ref{eq-chol-pol}) channel matrices, full channel matrix can be reconstructed using (\ref{eq-x-dec}). If spatially separated antennas are identical and both $\mathsf{R}$ and $\mathsf{L}$ polarizations are scattered in the same way we could potentially recover spatial correlation from uni-polarized measurements and polarimetric correlation -- from a single pair of dually polarized Tx and Rx antennas. This would allow to use partial measurements to predict capacity of a larger dimensions MIMO system.

Having full channel matrix $\mathbf{H}_{\times}$ built using (\ref{eq-x-dec}) decomposition into spatial and polarimetric matrices estimated using partial measurements, we can calculate ergodic capacity as an average of mutual information over all channel realizations for a given SNR $\gamma$ as
\begin{equation}
  C\left(\mathbf{H}_{\times}\right) = \mathbb{E}\left\{\log_{2}\det \left[\mathbf{I}_{N_{\rx}} + \frac{\gamma}{N_{\tx}} \mathbf{H}_{\times}\mathbf{H}^{\tsup{H}}_{\times} \right] \right\}.
\end{equation}
The capacity of reconstructed channel matrix can be compared against the capacity derived using original channel matrix.

\begin{figure*}
\centering
\subfigure[]{
   \includegraphics[width=0.47\textwidth]{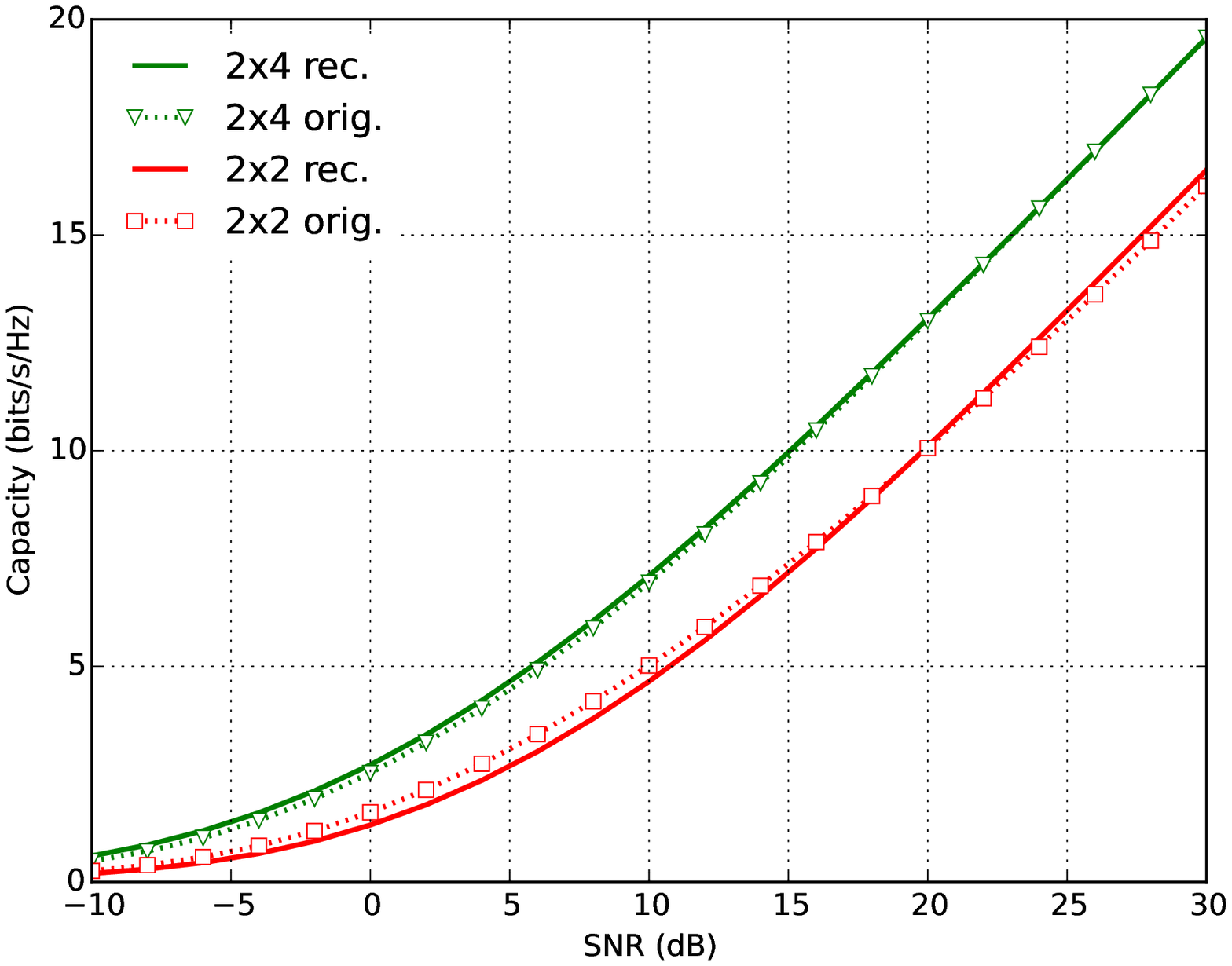}}
\subfigure[]{
   \includegraphics[width=0.47\textwidth]{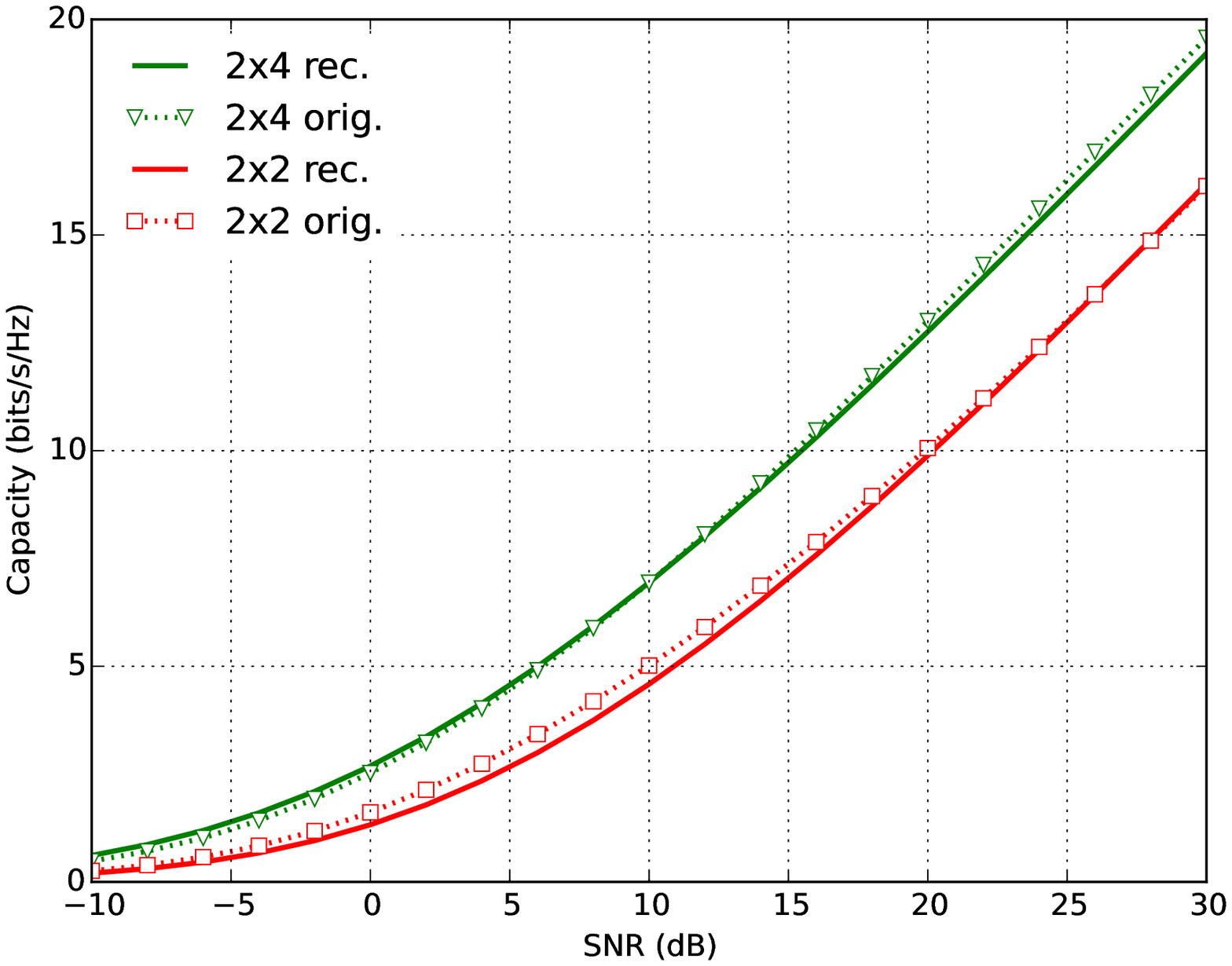}}
  \caption{Estimated ergodic capacity for MIMO $2\times2$ and $2\times4$ configurations: (a) for \emph{full} channel decomposition into spatial and polarimetric parts and (b) for channel reconstruction from the \emph{partial} spatial and polarimetric correlation matrices. Solid lines are for reconstructed capacity using (\ref{eq-x-dec}) decomposition, dotted lines -- capacity predicted using full-dimension correlation matrices obtained from measurements \cite{EkpePhD12}.}
\label{fig-capacity}
\end{figure*}

Example of land mobile satellite measurements in 2.5~GHz frequency band provided by Dr. Tim Brown from the University of Surrey, UK has been used to test the accuracy of reconstructing dual-polarized MIMO channel matrix from partial measurements. The measurement campaign has been carried out in suburban area of Guildford, UK \cite{EkpePhD12} with transmit antennas mounted on top of a tower and receiver antennas placed on a vehicle. In all cases described below, the transmitter has been equipped with two directional antennas one of which being right hand and and the other left hand circularly polarized. Two identical antennas have been used on the top of the vehicle with identical omni-directional radiation patterns, one for right hand circularly polarization and the other for left hand circularly polarization. The third antenna in the measurements has been different, a compact size antenna, therefore we used only the first two dual polarized Rx antennas resulting in maximal $2\times4$ MIMO configuration. We tested full channel (\ref{eq-x-dec}) matrix decomposition into spatial (\ref{eq-chol-sp}) and polarimetric (\ref{eq-chol-pol}) parts for single Rx antenna case ($2\times2$ MIMO configuration) and double Rx antenna case ($2\times4$ MIMO configuration) using averaging over both polarizations in (\ref{eq-R-sp}) and both antenna locations in (\ref{eq-R-pol}). Results of estimated capacity is shown in Fig.~\ref{fig-capacity} (a) comparing capacity obtained from original measurement data \cite{EkpePhD12} and reconstructed from (\ref{eq-x-dec}) decomposition. In a wide SNR range there is close similarity between original and reconstructed curves meaning the validity of full channel decomposition into spatial and polarimetric parts. Next the same $2\times2$ and $2\times4$ MIMO configurations have been reconstructed from minimal measured data set, i.e. using only $\mathsf{R}$ polarization in (\ref{eq-R-sp}) and only single Rx antenna location in (\ref{eq-R-pol}) reconstruction. The results of partial MIMO channel reconstruction displayed in Fig.~\ref{fig-capacity} (b) indicate similar capacity performance as with full channel information. The partial correlation information can be gathered using one measurement setup of two uni-polarized Rx antennas and another setup consisting of single Rx antenna with dual polarization measurements. Therefore full $2\times4$ MIMO system capacity in this case can be predicted having lower complexity MIMO measurement equipment, enabling estimation of spatial Rx correlation and polarization correlation separately during independent measurement runs.

\section{Conclusion}
MIMO channel matrix reconstruction from lower dimensional matrices are presented showing potential to use the method to approximate lower dimensional measurements for estimating performance of higher order MIMO configurations. Having the knowledge about internal structure of channel covariance matrix, such as Kronecker separability in spatial domain or decomposition into spatial and polarimetric correlation matrices, partial correlation matrices in one of these domains can be used to extrapolate MIMO system behavior in higher dimensions. Such extrapolation allows one to estimate channel statistics, capacity and power angular spectra in cluster scattering scenarios, including time-variant channel with Doppler shift. This would be useful in practical situations when only partial MIMO measurements are available. With proper assumptions these partial measurements could be used to predict higher order MIMO systems. It would enable estimation of channel capacity prior to installation of radio equipment when planning new wireless system deployments.

\begin{acknowledgements}
The author would like to thank Dr. Tim Brown from the University of Surrey, UK for providing MIMO land mobile satellite measurement data and Huawei Technologies (Vilnius) Ltd. for donating data center server used for numerical simulations.
\end{acknowledgements}


\begin{thebibliography}{10}
\providecommand{\url}[1]{{#1}}
\providecommand{\urlprefix}{URL }
\expandafter\ifx\csname urlstyle\endcsname\relax
  \providecommand{\doi}[1]{DOI \discretionary{}{}{}#1}\else
  \providecommand{\doi}{DOI \discretionary{}{}{}\begingroup
  \urlstyle{rm}\Url}\fi

\bibitem{DuelHallenChPred07}
Duel-Hallen A. (2007).
\newblock Fading {Channel} {Prediction} for {Mobile} {Radio} {Adaptive}
  {Transmission} {Systems}.
\newblock \emph{Proceedings of the IEEE} 95(12), 2299.
\newblock
  \href{http://dx.doi.org/10.1109/JPROC.2007.904443}{doi:10.1109/JPROC.2007.904443}

\bibitem{PhamChPred11}
Pham V.H., Wang X., Nadeau J. (2011).
\newblock Long {Term} {Cluster}-{Based} {Channel} {Envelope} and {Phase}
  {Prediction} for {Dynamic} {Link} {Adaptation}.
\newblock \emph{IEEE Communications Letters} 15(7), 713.
\newblock
  \href{http://dx.doi.org/10.1109/LCOMM.2011.051011.101684}{doi:10.1109/LCOMM.2011.051011.101684}

\bibitem{AdeogunChPred15}
Adeogun R., Teal P., Dmochowski P. (2015).
\newblock Extrapolation of {MIMO} {Mobile}-to-{Mobile} {Wireless} {Channels}
  {Using} {Parametric}-{Model}-{Based} {Prediction}.
\newblock \emph{IEEE Transactions on Vehicular Technology} 64(10), 4487.
\newblock
  \href{http://dx.doi.org/10.1109/TVT.2014.2366757}{doi:10.1109/TVT.2014.2366757}

\bibitem{KyostiEmulatingOTA12}
KyÃ¶sti P., JÃ¤msÃ¤ T., Nuutinen J.P. (2012).
\newblock Channel {Modelling} for {Multiprobe} {Over}-the-{Air} {MIMO}
  {Testing}.
\newblock \emph{International Journal of Antennas and Propagation} 2012, 1.
\newblock \href{http://dx.doi.org/10.1155/2012/615954}{doi:10.1155/2012/615954}

\bibitem{FanEmulatingOTA13}
Fan W., Lisbona X.C.B.d., Sun F., Nielsen J.Ã., Knudsen M.B., Pedersen G.F.
  (2013).
\newblock Emulating {Spatial} {Characteristics} of {MIMO} {Channels} for {OTA}
  {Testing}.
\newblock \emph{IEEE Transactions on Antennas and Propagation} 61(8), 4306.
\newblock
  \href{http://dx.doi.org/10.1109/TAP.2013.2261974}{doi:10.1109/TAP.2013.2261974}

\bibitem{FanReconstructionOTA13}
Fan W., Szini I., Nielsen J., Pedersen G. (2013).
\newblock Channel {Spatial} {Correlation} {Reconstruction} in {Flexible}
  {Multiprobe} {Setups}.
\newblock \emph{IEEE Antennas and Wireless Propagation Letters} 12, 1724.
\newblock
  \href{http://dx.doi.org/10.1109/LAWP.2014.2300096}{doi:10.1109/LAWP.2014.2300096}

\bibitem{WongSpatioTemporal06}
Wong I., Evans B. (2006).
\newblock Exploiting {Spatio}-{Temporal} {Correlations} in {MIMO} {Wireless}
  {Channel} {Prediction}.
  \href{http://dx.doi.org/10.1109/GLOCOM.2006.547}{doi:10.1109/GLOCOM.2006.547}

\bibitem{LiuSpatioTemporal14}
Liu L., Feng H., Yang T., Hu~B. (2014).
\newblock {MIMO}-{OFDM} {Wireless} {Channel} {Prediction} by {Exploiting}
  {Spatial}-{Temporal} {Correlation}.
\newblock \emph{IEEE Transactions on Wireless Communications} 13(1), 310.
\newblock
  \href{http://dx.doi.org/10.1109/TWC.2013.112613.130455}{doi:10.1109/TWC.2013.112613.130455}

\bibitem{KermoalKron02}
Kermoal J., Schumacher L., Pedersen K., Mogensen P., Frederiksen F. (2002).
\newblock A stochastic {MIMO} radio channel model with experimental validation.
\newblock \emph{IEEE Journal on Selected Areas in Communications} 20(6), 1211.
\newblock
  \href{http://dx.doi.org/10.1109/JSAC.2002.801223}{doi:10.1109/JSAC.2002.801223}

\bibitem{XiaoKron304}
Xiao C., Wu~J., Leong S.Y., Zheng Y., Letaief K. (2004).
\newblock A discrete-time model for triply selective {MIMO} {Rayleigh} fading
  channels.
\newblock \emph{IEEE Transactions on Wireless Communications} 3(5), 1678.
\newblock
  \href{http://dx.doi.org/10.1109/TWC.2004.833444}{doi:10.1109/TWC.2004.833444}

\bibitem{HanKron315}
Han B., Zheng Y. (2015).
\newblock Higher {Rank} {Principal} {Kronecker} {Model} for {Triply}
  {Selective} {Fading} {Channels} {With} {Experimental} {Validation}.
\newblock \emph{IEEE Transactions on Vehicular Technology} 64(5), 1654.
\newblock
  \href{http://dx.doi.org/10.1109/TVT.2014.2332518}{doi:10.1109/TVT.2014.2332518}

\bibitem{WernerCovEst08}
Werner K., Jansson M., Stoica P. (2008).
\newblock On {Estimation} of {Covariance} {Matrices} {With} {Kronecker}
  {Product} {Structure}.
\newblock \emph{IEEE Transactions on Signal Processing} 56(2), 478.
\newblock
  \href{http://dx.doi.org/10.1109/TSP.2007.907834}{doi:10.1109/TSP.2007.907834}

\bibitem{TsiligkaridisCovEst13}
Tsiligkaridis T., Hero A. (2013).
\newblock Covariance {Estimation} in {High} {Dimensions} {Via} {Kronecker}
  {Product} {Expansions}.
\newblock \emph{IEEE Transactions on Signal Processing} 61(21), 5347.
\newblock
  \href{http://dx.doi.org/10.1109/TSP.2013.2279355}{doi:10.1109/TSP.2013.2279355}

\bibitem{Oestges08}
Oestges C., Clerckx B., Guillaud M., Debbah M. (2008).
\newblock Dual-polarized wireless communications: from propagation models to
  system performance evaluation.
\newblock \emph{IEEE Transactions on Wireless Communications} 7(10), 4019.
\newblock
  \href{http://dx.doi.org/10.1109/T-WC.2008.070540}{doi:10.1109/T-WC.2008.070540}

\bibitem{ClerckxOestgesBook13}
Clerckx B., Oestges C., \emph{{MIMO} {Wireless} {Networks}, {Second} {Edition}:
  {Channels}, {Techniques} and {Standards} for {Multi}-{Antenna},
  {Multi}-{User} and {Multi}-{Cell} {Systems}}, 2nd edn. (Academic Press,
  Oxford; Waltham, MA, 2013)

\bibitem{HePolar16}
He~Y., Cheng X., Stuber G.L. (2016).
\newblock On polarization channel modeling.
\newblock \emph{IEEE Wireless Communications} 23(1), 80.
\newblock
  \href{http://dx.doi.org/10.1109/MWC.2016.7422409}{doi:10.1109/MWC.2016.7422409}

\bibitem{Liolis10}
Liolis K.P., Gomez-Vilardebo J., Casini E., Perez-Neira A.I. (2010).
\newblock Statistical {Modeling} of {Dual}-{Polarized} {MIMO} {Land} {Mobile}
  {Satellite} {Channels}.
\newblock \emph{IEEE Transactions on Communications} 58(11), 3077.
\newblock
  \href{http://dx.doi.org/10.1109/TCOMM.2010.091710.090507}{doi:10.1109/TCOMM.2010.091710.090507}

\bibitem{King12}
King P.R., Brown T.W.C., Kyrgiazos A., Evans B.G. (2012).
\newblock Empirical-{Stochastic} {LMS}-{MIMO} {Channel} {Model}
  {Implementation} and {Validation}.
\newblock \emph{IEEE Transactions on Antennas and Propagation} 60(2), 606.
\newblock
  \href{http://dx.doi.org/10.1109/TAP.2011.2173448}{doi:10.1109/TAP.2011.2173448}

\bibitem{CheffenaLMSPolar12}
Cheffena M., Fontan F.P., Lacoste F., Corbel E., Mametsa H.J., Carrie G.
  (2012).
\newblock Land {Mobile} {Satellite} {Dual} {Polarized} {MIMO} {Channel} {Along}
  {Roadside} {Trees}: {Modeling} and {Performance} {Evaluation}.
\newblock \emph{IEEE Transactions on Antennas and Propagation} 60(2), 597.
\newblock
  \href{http://dx.doi.org/10.1109/TAP.2011.2173447}{doi:10.1109/TAP.2011.2173447}

\bibitem{Loyka01}
Loyka S. (2001).
\newblock Channel capacity of {MIMO} architecture using the exponential
  correlation matrix.
\newblock \emph{IEEE Communications Letters} 5(9), 369.
\newblock \href{http://dx.doi.org/10.1109/4234.951380}{doi:10.1109/4234.951380}

\bibitem{Chizhik03}
Chizhik D., Ling J., Wolniansky P., Valenzuela R., Costa N., Huber K. (2003).
\newblock Multiple-input-multiple-output measurements and modeling in
  {Manhattan}.
\newblock \emph{IEEE Journal on Selected Areas in Communications} 21(3), 321.
\newblock
  \href{http://dx.doi.org/10.1109/JSAC.2003.809457}{doi:10.1109/JSAC.2003.809457}

\bibitem{GolubMatrix96}
Golub G.H., Van~Loan C.F., \emph{Matrix {Computations}}, 3rd edn. (Johns
  Hopkins University Press, Baltimore, MD, USA, 1996)

\bibitem{HerdinCMD05}
Herdin M., Czink N., Ozcelik H., Bonek E. (2005).
\newblock Correlation matrix distance, a meaningful measure for evaluation of
  non-stationary {MIMO} channels.
  \href{http://dx.doi.org/10.1109/VETECS.2005.1543265}{doi:10.1109/VETECS.2005.1543265}

\bibitem{WeichselbergerThesis03}
Weichselberger W., Spatial structure of multiple antenna radio channels.
\newblock {PhD} {Thesis}, Vienna University of Technology, Vienna, Austria
  (2003)

\bibitem{KunnariMIMOModel07}
Kunnari E., Iinatti J. (2007).
\newblock Stochastic modelling of {Rice} fading channels with temporal, spatial
  and spectral correlation.
\newblock \emph{IET Communications} 1(2), 215.
\newblock
  \href{http://dx.doi.org/10.1049/iet-com:20050654}{doi:10.1049/iet-com:20050654}

\bibitem{GhazalMIMONonstationary15}
Ghazal A., Wang C.X., Ai~B., Yuan D., Haas H. (2015).
\newblock A {Nonstationary} {Wideband} {MIMO} {Channel} {Model} for
  {High}-{Mobility} {Intelligent} {Transportation} {Systems}.
\newblock \emph{IEEE Transactions on Intelligent Transportation Systems} 16(2),
  885.
\newblock
  \href{http://dx.doi.org/10.1109/TITS.2014.2345956}{doi:10.1109/TITS.2014.2345956}

\bibitem{Buehrer02}
Buehrer R.M., in \emph{Wireless {Personal} {Communications}}, ed. by Tranter
  W.H., Woerner B.D., Reed J.H., Rappaport T.S., Robert M., no. 592 in The
  {International} {Series} in {Engineering} and {Computer} {Science} (Springer
  US, 2002), pp. 101--108.
  \href{http://dx.doi.org/10.1007/0-306-46986-3\_10}{doi:10.1007/0-306-46986-3\_10}

\bibitem{Bernado14}
Bernado L., Zemen T., Tufvesson F., Molisch A., Mecklenbrauker C. (2014).
\newblock Delay and {Doppler} {Spreads} of {Nonstationary} {Vehicular}
  {Channels} for {Safety}-{Relevant} {Scenarios}.
\newblock \emph{IEEE Transactions on Vehicular Technology} 63(1), 82.
\newblock
  \href{http://dx.doi.org/10.1109/TVT.2013.2271956}{doi:10.1109/TVT.2013.2271956}

\bibitem{MatzNonWSSUS05}
Matz G. (2005).
\newblock On non-{WSSUS} wireless fading channels.
\newblock \emph{IEEE Transactions on Wireless Communications} 4(5), 2465.
\newblock
  \href{http://dx.doi.org/10.1109/TWC.2005.853905}{doi:10.1109/TWC.2005.853905}

\bibitem{PaierWSSUS08}
Paier A., Zemen T., Bernado L., Matz G., Karedal J., Czink N., Dumard C.,
  Tufvesson F., Molisch A., Mecklenbrauker C. (2008).
\newblock Non-{WSSUS} vehicular channel characterization in highway and urban
  scenarios at 5.2 {GHz} using the local scattering function.
  \href{http://dx.doi.org/10.1109/WSA.2008.4475530}{doi:10.1109/WSA.2008.4475530}

\bibitem{MatzNonWSSUS03}
Matz G. (2003).
\newblock Doubly underspread non-{WSSUS} channels: analysis and estimation of
  channel statistics.
  \href{http://dx.doi.org/10.1109/SPAWC.2003.1318948}{doi:10.1109/SPAWC.2003.1318948}

\bibitem{Slepian78}
Slepian D. (1978).
\newblock Prolate {Spheroidal} {Wave} {Functions}, {Fourier} {Analysis}, and
  {Uncertainty} -- {V}: {The} {Discrete} {Case}.
\newblock \emph{Bell System Technical Journal} 57(5), 1371.
\newblock
  \href{http://dx.doi.org/10.1002/j.1538-7305.1978.tb02104.x}{doi:10.1002/j.1538-7305.1978.tb02104.x}

\bibitem{3GPP-TS36.104-16}
{3GPP}  (2016).
\newblock Evolved universal terrestrial radio access ({E}-{UTRA}); {Base}
  station ({BS}) radio transmission and reception.
\newblock Tech. Rep. TS 36.104 v13.5.0.
\newblock \urlprefix\url{http://www.3gpp.org/DynaReport/36104.htm}

\bibitem{Brown11}
Brown T.W.C., Kyrgiazos A. (2011).
\newblock On the small scale modelling aspects of dual circular polarised land
  mobile satellite {MIMO} channels in line of sight and in vehicles.
\newblock \urlprefix\url{http://ieeexplore.ieee.org/document/5782337/}

\bibitem{EkpePhD12}
Ekpe U., Modelling and {Measurement} {Analysis} of the {Satellite} {MIMO}
  {Radio} {Channel}.
\newblock {PhD} {Thesis}, University of Surrey (2012).
\newblock
  \urlprefix\url{http://ethos.bl.uk/OrderDetails.do?uin=uk.bl.ethos.553706}

\end{thebibliography}
\end{document}